\documentclass[a4paper,11pt]{article}

\usepackage{amsfonts, amsmath, amssymb}
\usepackage[utf8]{inputenc}
\usepackage[a4paper,top=3cm,bottom=2.5cm,left=2.5cm,right=2.5cm, bindingoffset=5mm]{geometry}
\usepackage{color}
\usepackage{booktabs}
\usepackage{subfigure}
\usepackage{graphicx}
\usepackage{float}
\usepackage{hyperref}
\hypersetup{citecolor=blue}
\usepackage{array}
\usepackage{colordvi}
\usepackage{multirow}
\usepackage{pifont} % \checkmark

\newcommand{\overbar}[1]{\mkern 1.5mu\overline{\mkern-1.5mu#1\mkern-1.5mu}\mkern 1.5mu}
\newcommand{\xmark}{\ding{55}}%

\begin{document}

\title{\bf Leptogenesis from low-energy CP violation in minimal left-right symmetric model}

\author{Xinyi Zhang$^{a}$\footnote{xzhang\_phy@pku.edu.cn}, Jiang-Hao Yu$^{b, c, d}$\footnote{corresponding author: jhyu@itp.ac.cn}, Bo-Qiang Ma$^{a,e,f}$ \\
}

\date{ 
$^a${\em \small School of Physics and State Key Laboratory of Nuclear Physics and Technology, \\ Peking University, Beijing 100871, China} \\
$^b${\em \small CAS Key Laboratory of Theoretical Physics, Institute of Theoretical Physics, Chinese Academy of Sciences,    \\ Beijing 100190, P. R. China}  \\
$^c${\em \small School of Physical Sciences, University of Chinese Academy of Sciences,   Beijing 100049, P.R. China}   \\ %[10mm]
$^d${\em \small School of Fundamental Physics and Mathematical Sciences, Hangzhou Institute for Advanced Study, \\ University of Chinese Academy of Sciences, Hangzhou 310024, China} \\
$^e${\em \small Collaborative Innovation Center of Quantum Matter, Beijing, China} \\
$^f${\em \small Center for High Energy Physics, Peking University, Beijing 100871, China}
}

\maketitle

\begin{abstract}
We perform a thermal unflavored leptogenesis analysis on minimal left-right symmetric models with discrete left-right symmetry identified as generalized parity or charge conjugation. When left-right symmetry is unbroken in the lepton Yukawa sector, the neutrino Dirac coupling matrix is completely determined by neutrino masses and mixing angles, allowing CP violation needed to generate leptogenesis totally resides in the low-energy sector. 
With two lepton asymmetry generation ways, both type I and mixed type I$+$II neutrino mass generation mechanisms are considered. After solving the Boltzmann equations numerically, we find that the low-energy CP phases in the lepton mixing matrix can successfully produce the observed baryon asymmetry, and in some cases, the Dirac CP phase can be the only source of CP violation. Finally, we discuss the interplay among low-energy CP phase measurements, leptogenesis, and neutrinoless double beta decay. We show that the viable models for successful leptogenesis can be probed in next-generation neutrinoless double-beta decay experiments.
\end{abstract}

\tableofcontents

\section{Introduction}

The baryon asymmetry of the Universe (BAU) remains mysterious in the development of modern physics. Sakharov proposes~\cite{Sakharov:1967dj} that there are three conditions to be satisfied to have a dynamically generated baryon asymmetry: the baryon number violation, C and CP violation, and departure from thermal equilibrium. Among various attempts to explain BAU dynamically, leptogenesis~\cite{Fukugita:1986hr} is quite intriguing because it not only satisfies the Sakharov conditions naturally but also explains the lightness of neutrino mass through the seesaw mechanism. It is even more interesting when the CP violation in the neutrino sector is the only CP-violating source needed for generating BAU. When this happens, physics becomes so neat: what we need to explain things at low energy also tells us about high energy and the early Universe. No additional source of CP violation at high scale is needed in explaining BAU.

In typical seesaw models, there is no direct connection between the low-energy CP violation and high scale leptogenesis. In the type I seesaw, the neutrino Dirac coupling in the high-scale Lagrangian is important in generating the light neutrino mass at a low scale and makes a dominant contribution to the CP asymmetry generated by heavy neutrino decay. After electroweak symmetry breaking, the neutrino Dirac coupling becomes a $3\times 3$ complex matrix whose entities cannot be totally determined by the low-energy  neutrino parameters such as neutrino masses, mixing angles, and CP phases. One can see this by referring to the Casas-Ibarra parameterization~\cite{Casas:2001sr}, with no further assumptions in type I seesaw, the neutrino Dirac coupling matrix $M_\mathrm{D}$ can be expressed with neutrino parameters plus an arbitrary orthogonal matrix $R$ as
\begin{align}
\tilde{M}_\mathrm{D} = i D_N^{1/2} R D_\nu^{1/2} V_\mathrm{L}^\dagger, \label{eq:CI}
\end{align}
where $D_\nu$ is the diagonal neutrino mass matrix which is related to the non-diagonal one by $M_\nu=V_\mathrm{L}^* D_\nu^{} V_\mathrm{L}^\dagger$, $V_\mathrm{L}$ is the lepton mixing matrix, $D_N$ is the diagonal heavy neutrino mass matrix, and $\tilde{M}_\mathrm{D} $ is the neutrino Dirac coupling matrix in right-handed neutrino mass basis. A similar parameterization is proposed in Ref.~\cite{Akhmedov:2008tb} for type I$+$II seesaw, which also contains an arbitrary complex orthogonal matrix in full analogy with $R$.

To make the connection between the low-energy CP violation and leptogenesis, there are many studies with certain assumptions~\cite{Branco:2001pq,Branco:2002xf,Endoh:2002wm,Frampton:2002qc,Pascoli:2006ie,Pascoli:2006ci,Molinaro:2008rg}. By assuming a CP-conserving $R$ matrix, Ref.~\cite{Moffat:2018smo} shows there are viable parameter space across seven orders of magnitude $10^6~\mathrm{GeV} < T < 10^{13}~\mathrm{GeV}$; At high energy $T \gg 10^{12}$ GeV, leptogenesis is still sensitive to low-energy phases. Furthermore, flavor symmetries are widely utilized to generate patterns of neutrino Dirac coupling and thus are able to fix the orthogonal matrix $R$~\cite{Hagedorn:2009jy,Meroni:2012ze,Karmakar:2014dva,Gehrlein:2015dxa,Ishihara:2015uua,Li:2017zmk}. 
We take a quite different approach in this work by utilizing the left-right symmetry in the leptonic Yukawa sector to remove the arbitrary orthogonal matrix $R$ completely.

Minimal left-right symmetric model (MLRSM)~\cite{Pati:1974yy,Mohapatra:1974gc,Senjanovic:1975rk,Senjanovic:1978ev}, provides a natural framework for neutrino mass generation and leptogenesis, with rich phenomena~\cite{Joshipura:2001ya,Rodejohann:2002mh,Babu:2005bh,Akhmedov:2006yp,Chao:2007rm,Hallgren:2007nq,Abada:2008gs,Dev:2019rxh,Rink:2020uvt}. The simultaneous presence of right-handed neutrino and left-handed triplet scalar enables type I~\cite{Minkowski:1977sc,Yanagida:1980xy,Mohapatra:1980yp} and/or type II~\cite{Magg:1980ut,Schechter:1980gr,Wetterich:1981bx,Lazarides:1980nt} seesaw for neutrino mass generation. Furthermore, they also provides two possible ways of lepton asymmetry generation by acting as the decaying particle to generate the CP asymmetry separately. The various leptogenesis scenarios in left-right symmetric theories are discussed in Ref.~\cite{Hambye:2003ka}.

There are many successful realizations of leptogenesis in left-right symmetric theories with different focuses and different measures to constrain the free parameters as in the neutrino Dirac coupling matrix (see e.g., Refs.~\cite{Joshipura:2001ya,Rodejohann:2002mh,Babu:2005bh,Akhmedov:2006yp,Chao:2007rm,Hallgren:2007nq,Abada:2008gs,Rink:2020uvt}). For example, 
Refs.~\cite{Joshipura:2001ya,Rodejohann:2002mh,Abada:2008gs,Rink:2020uvt} adopt a SO(10) GUT inspired relation $M_\mathrm{D} = M_\mathrm{u}$, where $M_\mathrm{u}$ is the up-type quark mass matrix. Refs.~\cite{Joshipura:2001ya,Rodejohann:2002mh,Rink:2020uvt} relate the lepton asymmetry to low-energy parameters by further assuming that only top quark mass contributes and working with type II seesaw dominated light neutrino mass. Ref.~\cite{Babu:2005bh} expresses the lepton asymmetry in terms of nine low-energy measurable parameters, by assuming low-energy supersymmetry and taking that $M_\mathrm{D} \propto M_l$, where $M_l$ is the charged lepton mass matrix. Ref.~\cite{Akhmedov:2006yp} addresses the eight-fold degeneracy of the heavy neutrino mass matrix~\cite{Akhmedov:2005np} and takes $M_\mathrm{D}$ to be symmetric. Ref.~\cite{Chao:2007rm} realizes type II seesaw leptogenesis in left-right symmetric models with the Higgs triplet Yukawa coupling matrix taking the Friedberg-Lee texture~\cite{Friedberg:2006it}. Ref.~\cite{Hallgren:2007nq} investigates triplet leptogenesis in left-right symmetric theories with symmetric $M_\mathrm{D}$. Ref.~\cite{Abada:2008gs} studies left-right symmetric seesaw in supersymmetric SO(10) models where $M_\mathrm{D} =M_\mathrm{u}$ holds.

In this work, we investigate leptogenesis in the MLRSM with unbroken left-right symmetry identified as the parity or charge conjugate symmetry~\cite{Nemevsek:2012iq,Senjanovic:2016vxw,Senjanovic:2018xtu,Senjanovic:2019moe}. Recently, it has been pointed out~\cite{Nemevsek:2012iq,Senjanovic:2016vxw,Senjanovic:2018xtu,Senjanovic:2019moe} that in MLRSM with unbroken left-right symmetry in lepton Yukawa sector, light and heavy neutrino masses and mixings are sufficient to determine the neutrino Dirac matrix. In previous literatures on leptogenesis in left-right symmetric models, usually some assumptions are utilized to fix the orthogonal matrix $R$, e.g., GUT-inspired mass matrix relation~\cite{Joshipura:2001ya,Rodejohann:2002mh,Abada:2008gs,Rink:2020uvt}, proportionality with the charged lepton mass~\cite{Babu:2005bh}, being symmetric~\cite{Akhmedov:2006yp,Hallgren:2007nq}, or flavor textures~\cite{Chao:2007rm}. We realize that the left-right symmetry, as the result of the MLRSM setup, provides a new form of the constrained $R$ matrix, which removes the $R$ matrix ambiguity with no need to make some assumptions. This allows us to probe the link between low-energy CP violation and BAU. We present here in a neat and special case where the light and heavy neutrino mixing coincidence (up to a conjugation), leaving the CP violation needed to generate the baryon asymmetry of the Universe entirely in the light neutrino sector.
The coincidence of light and heavy neutrino mixing is not mandatory but only serves as a reasonable choice that resembles the quark sector situation~\cite{Senjanovic:2014pva}. This choice greatly increases the model's predicting power and avoids complications caused by the heavy neutrino mixing. Since the heavy neutrino can provide extra CP-violating sources, we are optimistic that given positive results in this study, enough BAU can be generated with a more general heavy neutrino mixing.

%In this work, we perform a leptogenesis analysis given the neutrino Dirac coupling structure in MLRSM. This structure leaves no ambiguity as the $R$ matrix (or its equivalent in mixed type I + II seesaw) is now fixed and allows a single flavor analysis. Working in MLRSM, we consider both charge conjugation and parity as the left-right symmetry. With unbroken left-right symmetry in the lepton Yukawa sector, the CP violation needed in leptogenesis totally resides in the low energy sector. We also take into account different neutrino mass generation and CP asymmetry generation mechanisms. The Boltzmann equations are solved numerically, and we find viable parameter space exists.

The paper is organized as follows. In Sec.~\ref{sec:Dirac}, we review how the neutrino Dirac coupling matrix is completely determined by the light and heavy neutrino masses and mixings in MLRSM. We consider both charge conjugation and generalized parity as left-right symmetry. In Sec.~\ref{sec:lep}, we present the CP asymmetries and the Boltzmann equations for analyzing leptogenesis. The numerical results are collected in Sec.~\ref{sec:result}. We discuss the interplay among low-energy CP violation, leptogenesis and neutrinoless double beta decay in Sec.~\ref{sec:interplay} and conclude in Sec.~\ref{sec:conclusion}. In Appendices, we collect supporting materials.

\section{Determine neutrino Dirac coupling in MLRSM}\label{sec:Dirac}

MLRSM~\cite{Pati:1974yy,Mohapatra:1974gc,Senjanovic:1975rk,Senjanovic:1978ev} is based on the gauge group $SU(2)_\mathrm{L} \times SU(2)_\mathrm{R} \times U(1)_{\mathrm{B}-\mathrm{L}}$ plus a discrete left-right (LR) symmetry which leads to the equality of the gauge coupling, $g_\mathrm{L} =g_\mathrm{R} =g$. This model has been widely studied, so we only show the Lagrangian relevant for the following discussion here
\begin{align}
\mathcal{L} \supset -\bar{l}_\mathrm{L}^{} \left(Y_1 \Phi_1 - Y_2^{} \Phi_2^* \right) l_\mathrm{R}^{} - \frac{1}{2} \left(l_\mathrm{L}^T\mathrm{C} Y_\mathrm{L}^{} i \sigma_2^{} \Delta_\mathrm{L}^{} l_\mathrm{L}^{}
 + l_\mathrm{R}^T  \mathrm{C}  Y_\mathrm{R}^{} i \sigma_2^{} \Delta_\mathrm{R}^{} l_\mathrm{R}^{}\right) -\lambda_{ij} \mathrm{Tr} \left( \Delta_\mathrm{R}^\dagger \Phi_i^{} \Delta_\mathrm{L}^{} \Phi_j^\dagger \right)+ \mathrm{h.c.},\label{eq:Lag}
\end{align}
where 
\begin{align}
l_{\mathrm{L,R}}=\left(\begin{array}{c}
\nu\\ e\\
\end{array}\right)_\mathrm{L,R},~~
\Delta_\mathrm{L,R}=\left(\begin{array}{cc}
\delta^+/\sqrt{2}  & \delta^{++}\\
\delta^0 & -\delta^+/\sqrt{2}\\
\end{array}\right)_\mathrm{L,R},~~
\Phi_1= \left( \begin{array}{cc}
\phi_1^0 & \phi_2^+\\
\phi_1^- & -\phi_2^0\\
\end{array}\right),~~
\Phi_2=\sigma_2^{} \Phi_1^* \sigma_2^{}.
\end{align}
We also show their representations in MLRSM gauge groups in Table~\ref{tab:assignment}. For more details, we refer to Refs.~\cite{Nemevsek:2012iq,Senjanovic:2016vxw,Senjanovic:2018xtu,Senjanovic:2019moe,Hsieh:2010zr, Cao:2012ng}.

\begin{table}[t!]
 \caption{\label{tab:assignment} Field representations under the gauge groups in MLRSM.}\vspace{0.12cm}
 \centering
  \begin{tabular}{c|c|c|c}
   \toprule \hline
   & $SU(2)_\mathrm{L}$ & $SU(2)_\mathrm{R}$ & $U(1)_{\mathrm{B}-\mathrm{L}}$\\ \hline
   $l_\mathrm{L}$  &  2 & 1 & -1 \\
   $l_\mathrm{R}$  & 1 & 2 & -1 \\
   $\Delta_\mathrm{L}$ & 3 & 1 & 2\\
    $\Delta_\mathrm{R}$ & 1 & 3 & 2\\
   $\Phi_1$ & 2 & 2 & 0 \\
   $\Phi_2$ & 2 & 2 & 0 \\
   \hline
  \bottomrule
\end{tabular}
\end{table}

Typically the symmetry breakings happen in several steps~\cite{Hsieh:2010zr,Cao:2012ng}: first the right-handed triplet scalar $\Delta_\mathrm{R}$ obtains vacuum expectation value (vev) at a high scale, and then the electroweak symmetry breaks, which triggers the induced symmetry breaking for $\Delta_\mathrm{L}$.  
These symmetry breakings usually mix the two Higgs doublets  $\phi_i=(\phi_i^0,\phi_i^-)^T$($i = 1, 2$), and the mass eigenstates are $ H=(v_1 \phi_1 + v_2 \phi_2)/\sqrt{v_1^2+v_2^2}$ corresponding to the standard model Higgs doublet and ${ H^\prime}= (v_2 \phi_1 - v_1 \phi_2)/\sqrt{v_1^2+v_2^2}$ the new Higgs doublet. 
We consider that both $\Delta_\mathrm{R}$ and $ H^\prime$ are much heavier than $\Delta_\mathrm{L}$, which could be naturally realized by assuming all the coupling coefficients in the scalar potential have a similar size. For the expression of the full scalar potential, we refer to, e.g., Ref.~\cite{Deshpande:1990ip}. After integrating out $\Delta_\mathrm{R}$ and $ H^\prime$, which have masses proportional to $v_\mathrm{R}$ and thus are decoupled~\cite{Hambye:2003ka,Senjanovic:2018xtu}, we arrive at the low-energy  scalar potential, in which the trilinear scalar coupling $\Delta_\mathrm{L} \Phi \Phi$ is relevant to the lepton asymmetry considered below.

After $\Delta_\mathrm{R}$ acquires a vev $\langle \Delta_\mathrm{R} \rangle =v_\mathrm{R}$ and the Higss bidoublet gets vev $\langle \phi_i^0 \rangle = v_i$ ($i = 1, 2$), one arrives at the effective trilinear scalar coupling term $\mu H^T i \sigma_2 \Delta_\mathrm{L} H$ with $\mu$  being determined from the scalar quartic term in Eq.~(\ref{eq:Lag}) as 
\begin{align}
\displaystyle \mu = \frac{(\lambda_{11}+\lambda_{22}) v_1 v_2 + \lambda_{12} v_2^2 +\lambda_{21} v_1^2}{v_1^2+v_2^2} v_\mathrm{R}.
\end{align}
It is also related to the left-handed triplet vev $\langle \Delta_\mathrm{L} \rangle =v_\mathrm{L}$ and the left-handed triplet mass $m_\Delta$ as $\displaystyle \mu = v_\mathrm{L} m_\Delta^2/v^2$ ~\cite{Du:2018eaw}.

In the following, we briefly review the findings in Refs.~\cite{Nemevsek:2012iq,Senjanovic:2016vxw,Senjanovic:2018xtu,Senjanovic:2019moe} and show that the Dirac coupling in the MLRSM can be completely determined by light and heavy neutrino masses and mixings in both charge conjugation ($\mathcal{C}$) and generalized parity ($\mathcal{P}$) as the LR symmetry cases.

\subsection{$\mathcal{C}$ as the LR symmetry}

When $\mathcal{C}$ is the LR symmetry, the fields transform under $\mathcal{C}$ as 
\begin{align}
l_\mathrm{L}^{} \leftrightarrow l_\mathrm{R}^\mathrm{c},~~\Delta_\mathrm{L}^{} \leftrightarrow \Delta_\mathrm{R}^*,~~\Phi^{}_{} \leftrightarrow \Phi^T_{},
\end{align}
then the relations of the coupling matrices can be derived
\begin{align}
Y_{1,2}^{}=Y_{1,2}^T,~~Y_\mathrm{L}^{}=Y_\mathrm{R}^* \equiv Y_\mathrm{T},\label{eq:Ysymmetric}
\end{align}
which leads to a symmetric neutrino Dirac coupling matrix, i.e., $M_\mathrm{D}^{} = M_\mathrm{D}^T$.

The effective light neutrino mass is 
\begin{align}
M_\nu = \frac{v_\mathrm{L}}{v_\mathrm{R}}  M_N  - M_\mathrm{D}^T\frac{1}{M_N}M_\mathrm{D}^{} \equiv M_\nu^\mathrm{II} + M_\nu^\mathrm{I},
\end{align}
where we introduce $N_\mathrm{L} \equiv C \bar{\nu}_\mathrm{R}$ such that $M_N =v_\mathrm{R}^{} Y_\mathrm{R}^*$.
With $M_\mathrm{D}$ being symmetric, one can get 
\begin{align}
M_\mathrm{D}=M_N \sqrt{\frac{v_\mathrm{L}}{v_\mathrm{R}}-\frac{1}{M_N}M_\nu}.\label{eq:MD_C}
\end{align}
We see that given the light and heavy neutrino masses and mixings, the neutrino Dirac coupling matrix is totally fixed. It is useful to see this in the type I seesaw limit. In $M_\nu \simeq M_\nu^\mathrm{I}$ case, one has
\begin{align}
M_\mathrm{D}= i M_N \sqrt{\frac{1}{M_N}M_\nu}.
\end{align}
Converting to the right-handed neutrino mass basis and comparing Eq.(\ref{eq:CI}) with $\tilde{M}_\mathrm{D}$ in our case, we find %\footnote{This result is different from Eq.(9) in Ref.~\cite{Nemevsek:2012iq}, where $M_\mathrm{D}$ in weak basis is mistakenly compared with the mass basis $\tilde{M}_\mathrm{D} $ in Casas-Ibarra parameterization. }
\begin{align}
R=V_\mathrm{L}^*. \label{eq:RinC}
\end{align}
In this particular case, $R$ matrix coincides with the light neutrino mixing matrix up to a charge conjugation. Nevertheless, as it only contains phases from the low-energy sector, it provides no additional CP-violating source.

\subsection{$\mathcal{P}$ as the LR symmetry}

Under generalized parity transformation, the fields behave as
\begin{align}
l_\mathrm{L} \leftrightarrow l_\mathrm{R},~~\Delta_\mathrm{L} \leftrightarrow \Delta_\mathrm{R},~~\Phi^{} \leftrightarrow \Phi^\dagger,
\end{align}
which leads to
\begin{align}
Y_{1,2}^{}=Y_{1,2}^\dagger,~~Y_\mathrm{L}=Y_\mathrm{R} \equiv Y_\mathrm{T},\label{eq:Yhermitian}
\end{align}
which results in $M_\mathrm{D}=M_\mathrm{D}^\dagger$ in the case of unbroken parity. More details can be found in Refs.~\cite{Senjanovic:2016vxw,Senjanovic:2018xtu,Senjanovic:2019moe}.

The effective neutrino mass is now
\begin{align}
M_\nu = \frac{v_\mathrm{L}}{v_\mathrm{R}}  M_N^* - M_\mathrm{D}^T \frac{1}{M_N}M_\mathrm{D} = M_\nu^\mathrm{II} + M_\nu^\mathrm{I},\label{eq:Mnu}
\end{align}
from which and using $M_\mathrm{D}=M_\mathrm{D}^\dagger$, one can introduce a hermitian matrix $H$ and define
\begin{align}
H^{} H^T \equiv \frac{v_\mathrm{L}^*}{v_\mathrm{R}} - \frac{1}{\sqrt{M_N}} M_\nu^* \frac{1}{\sqrt{M_N}} . \label{eq:HHT}
\end{align}
$M_\mathrm{D}$ can be expressed in terms of $H$ as
\begin{align}
M_\mathrm{D} =\sqrt{M_N} H \sqrt{M_N^*}. \label{eq:MDH}
\end{align}
Working in the type I seesaw limit $M_\nu \simeq M_\nu^\mathrm{I}$ and in right-handed neutrino mass basis, Eq.(\ref{eq:HHT}) can be written as
\begin{align}
H^{} H^T \simeq  - D_N^{-1/2} M_\nu^* D_N^{-1/2}
= - D_N^{-1/2} V_\mathrm{L}^{} D_\nu^{} V_\mathrm{L}^T  D_N^{-1/2}.
\end{align}
By identifying $H= i D_N^{-1/2} V_\mathrm{L} D_\nu^{1/2}$,  we get
\begin{align}
\tilde{M}_\mathrm{D} = D_N^{1/2} H D_N^{1/2} = i V_\mathrm{L}^{} D_\nu^{1/2}  D_N^{1/2}. \label{eq:MDtypeI}
\end{align}
Comparing Eq.(\ref{eq:MDtypeI}) with the Dirac Yukawa term written in the Casas-Ibarra parameterization in Eq.~(\ref{eq:CI}), we now have
\begin{align}
R = D_N^{-1/2} V_\mathrm{L}^{} D_\nu^{1/2} D_N^{1/2} V_\mathrm{L}^{} D_\nu^{-1/2}, \label{eq:RinP}
\end{align}
which is again determined by the light and heavy neutrino mass information, leaving no ambiguity in neutrino Dirac mass term.

\section{Leptogenesis in MLRSM}\label{sec:lep}

When leptogenesis in MLRSM is considered, depending on the mechanisms of light neutrino mass generation and lepton asymmetry generation, there are in general four cases~\cite{Hambye:2003ka}:
\begin{itemize}
\item  Case 1: type I contribution dominates light neutrino mass, heavy neutrinos decay;
\item  Case 2: type II contribution dominates light neutrino mass, heavy neutrinos decay; 
\item  Case 3: type I contribution dominates light neutrino mass, left-handed triplet scalars decay; 
\item  Case 4: type II contribution dominates light neutrino mass, left-handed triplet scalars decay.
 \end{itemize}
Notice that we do not consider the case that neutrino masses are generated by a cancellation among the type II and type I term to avoid a fine tuning of the parameters. There are intermediate cases that either type I and type II mass contributions are comparable, or lepton asymmetries generated by both heavy neutrinos and left-handed triplet scalars are relevant. Our discussions on the four limiting cases can be applied to the intermediate-mass contribution cases, but not the latter. We consider the limiting cases that either heavy neutrinos or left-handed triplet scalars decay matters, under the assumption that their masses are very hierarchical. Thus the lighter one will effectively wash out all the CP asymmetries generated by the heavier one. In the temperature range of leptogenesis, the heavy one effectively decouples, and only the light one is responsible for the final baryon asymmetry. In cases that the heavy neutrino mass is comparable with the left-handed triplet mass, the Boltzmann equations for right-handed neutrinos and left-handed triplets are coupled, and all the three CP asymmetries defined in the following will contribute to the same lepton number density. In such a case, it is hard to estimate the results before solving the Boltzmann system. For simplicity, we follow Ref.~\cite{Hambye:2003ka} and consider only the four limiting cases. We work in the basis that the heavy neutrino mass matrix is real and diagonal, which corresponds to change $M_\mathrm{D}$ into $\tilde{M}_\mathrm{D} = V_\mathrm{R} M_\mathrm{D} $ \footnote{Note $\overbar{\nu}_\mathrm{R} M_\mathrm{D} \nu_\mathrm{L} = \overbar{\tilde{\nu}}_\mathrm{R} \tilde{M}_\mathrm{D} \nu_\mathrm{L}$. Since $\tilde{\nu}_\mathrm{R}=V_\mathrm{R} \nu_\mathrm{R}$, we get $\tilde{M}_\mathrm{D}=V_\mathrm{R} M_\mathrm{D}  $. } accordingly. For the sake of convenience, we drop the tilde notation for quantities in the transformed basis, and it should be understood without confusion.
In this section, we set up the basic ingredients for leptogenesis and present our results in the next section.

\subsection{The CP asymmetries}

We set the temperature that leptogenesis takes place to be beyond $10^{12}$ GeV, so that we can work with the single-flavour approximation. The lepton number asymmetry can be generated either by the lightest heavy neutrino or the left-handed triplet scalar, depending on their relative masses.

\begin{figure}[t!]
\centering
\includegraphics[width=0.25\textwidth]{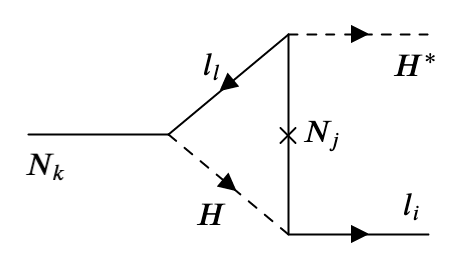}\hspace{1cm}
\includegraphics[width=0.25\textwidth]{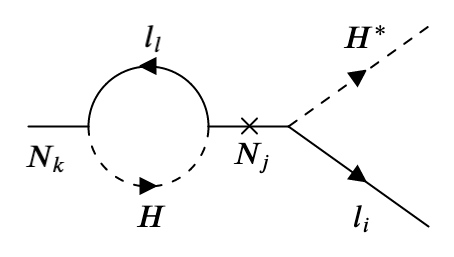}\hspace{1cm}
\includegraphics[width=0.25\textwidth]{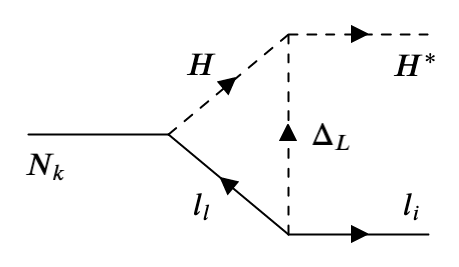}
\caption{One loop diagrams for the heavy neutrino decay.
}
\label{fig:Ndecay}
\end{figure}

\begin{figure}[t!]
\centering
\includegraphics[width=0.3\textwidth]{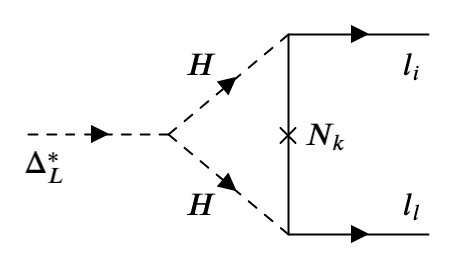}
\caption{One loop diagram for the left-handed triplet scalar decay.
}
\label{fig:DeltaDecay}
\end{figure}

The CP asymmetry generated by a heavy neutrino decay is
\begin{align}
\epsilon_{N_k}=\displaystyle\sum_i \frac{\Gamma(N_k\rightarrow l_i H^*)- \Gamma(N_k\rightarrow \bar{l}_i H)}
{\Gamma(N_k\rightarrow l_i H^*)+ \Gamma(N_k\rightarrow \bar{l}_i H)},
\end{align}
which is generated by the interference between the tree-level decay amplitude with one-loop amplitudes as shown in Fig.~\ref{fig:Ndecay}. 
The CP asymmetry generated by interference with the first two one-loop amplitudes is the usual type I seesaw asymmetry
\begin{align}
\epsilon_{N_k} =\displaystyle\frac{1}{8\pi}\sum_j \frac{\mathrm{Im} (Y_NY_N^\dagger)_{kj}^2}{\sum_i |(Y_N)_{ki}|^2} \sqrt{x_j} 
\left[ 1-(1+x_j) \mathrm{ln}(1+\frac{1}{x_j})+\frac{1}{1-x_j} \right],\label{eq:epsilonN}
\end{align}
where $Y_N = M_\mathrm{D}/v,~x_j = m_{N_j}^2/m_{N_k}^2$.
The third diagram contributes~\cite{Hambye:2003ka}
\begin{align}
\epsilon_{N_k}^\Delta =\displaystyle -\frac{1}{2 \pi} 
\frac{\sum_{il} \mathrm{Im}\left[(Y_N)_{ki} (Y_N)_{kl} (Y_\mathrm{T}^*)_{il} \mu\right]}{\sum_i |(Y_N)_{ki}|^2 m_{N_k}} 
\left[ 1- \frac{m_\Delta^2}{m_{N_k}^2} \mathrm{ln}\left(1+\frac{m_{N_k}^2}{m_\Delta^2}\right)  \right],\label{eq:epsilonND}
\end{align}
where $Y_\mathrm{T} = M_N/v_\mathrm{R}$ when $\mathcal{C}$ is the LR symmetry, and $Y_\mathrm{T} = M_N^*/v_\mathrm{R}$ when $\mathcal{P}$ is the LR symmetry. 

In the usual case, single-flavor leptogenesis is independent of the lepton mixing matrix, as can be seen from
\begin{align}
Y_N^{} Y_N^\dagger = \displaystyle \frac{1}{v^2} M_\mathrm{D}^{} M_\mathrm{D}^\dagger 
= \displaystyle \frac{1}{v^2} D_N^{1/2} R _{}^{} D_\nu R_{}^\dagger D_N^{1/2},
\end{align}
where in the equality we use the Casas-Ibarra parameterization for $M_\mathrm{D}$ as shown in Eq.(\ref{eq:CI}).  With no knowledge on the $R$ matrix, this expression has no dependence on the lepton mixing matrix. However, with the $R$ matrix  in Eq.(\ref{eq:RinC}) and Eq.(\ref{eq:RinP}) determined from the LR symmetry, one can see that $Y_N^{} Y_N^\dagger$ are 
\begin{align}
Y_N^{} Y_N^\dagger &= \displaystyle \frac{1}{v^2} D_N^{1/2} V_\mathrm{L}^* D_\nu^{} V_\mathrm{L}^T D_N^{1/2},~(\mathcal{C})\\
Y_N^{} Y_N^\dagger&= \displaystyle \frac{1}{v^2}  V_\mathrm{L}^{} D_\nu^{1/2} D_N^{} D_\nu^{1/2} V_\mathrm{L}^\dagger,~(\mathcal{P})
\end{align}
where $\mathcal{C}$ and $\mathcal{P}$ mark the LR symmetry. We see that $Y_N^{} Y_N^\dagger$ are dependent on the lepton mixing, and so are the CP asymmetries. As a result, we can investigate the role of the low-energy CP violation in the single-flavor regime of leptogenesis.

The CP asymmetry generated by the left-handed triplet decay is produced by the interference of the tree level amplitude with the one-loop diagram shown in Fig.~\ref{fig:DeltaDecay}. The asymmetry is 
\begin{align}
\epsilon_\Delta &= 2 \sum_{i,l} \displaystyle\frac{\Gamma(\Delta_\mathrm{L}^*\rightarrow l_i l_l)- \Gamma(\Delta_\mathrm{L} \rightarrow \bar{l}_i \bar{l}_l)}{\Gamma(\Delta_\mathrm{L}^*\rightarrow l_i l_l)+ \Gamma(\Delta_\mathrm{L} \rightarrow \bar{l}_i \bar{l}_l)} \\
&=\displaystyle\frac{1}{8\pi} \sum_k m_{N_k} \frac{\sum_{il} \mathrm{Im} \left[ (Y_N^*)_{ki} (Y_N^*)_{kl} (Y_\mathrm{T})_{il} \mu^* \right] }{\sum_{ij}|(Y_\mathrm{T})_{ij}|^2 m_\Delta^2 + |\mu|^2 }  \mathrm{ln}\left(1+\frac{m_\Delta^2}{m_{N_k}^2}\right). \label{eq:epsilonD}
\end{align}

\subsection{The Boltzmann equations}

To quantitatively examine the dynamics generating the final baryon asymmetry, we need the transport equations for relevant species.  

In the cases that the heavy neutrinos decay, when their masses are not very hierarchical, we take into consideration all the three right-handed neutrinos decay. The set of Boltzmann equations reads
\begin{align}
\frac{dY_{N_i}}{dz} &= \frac{-z}{s H} 
\left( \gamma_\mathrm{D}^i + \gamma_{\mathrm{S},\Delta \mathrm{L}=1}^i \right) \left( \frac{Y_{N_i}}{Y_{N_i}^{\rm eq}} -1 \right); \\
\frac{dY_\Delta}{dz} &=  \frac{-z}{s H} 
\left[ \sum_{i=1}^3 \epsilon_i \left( \gamma_\mathrm{D}^i + \gamma_{\mathrm{S},\Delta \mathrm{L}=1}^i \right) \left( \frac{Y_{N_i}}{Y_{N_i}^{\rm eq}} -1 \right)
- \left(\frac{ \gamma_\mathrm{D}^i}{2} + \gamma_{\mathrm{W},\Delta \mathrm{L}=1}^i \right) \frac{Y_\Delta}{Y_l^{\rm eq}}
\right],
\end{align}
where $Y_{N_i}$ is the $N_i$ abundance ($Y_{N_i}^{\rm eq}$ is the equilibrium abundance), $\displaystyle Y_\Delta \equiv Y_{\Delta \mathrm{B}}/3 -Y_{\Delta \mathrm{L}}$ and $\epsilon_i \equiv \epsilon_{N_i} + \epsilon_{N_i}^\Delta$. Note also $\displaystyle z =m_{N_1}/T$ with $T$ being the temperature of the thermal bath, $s$ is the entropy density and $H(T)$ is the Hubble expansion rate of the Universe. The rate density $\gamma_\mathrm{D}^i$ is the thermally averaged total decay rate of $N_i$ into the SM lepton and Higgs doublets, $\gamma_{\mathrm{S},\Delta \mathrm{L}=1}^i $ is the $\Delta \mathrm{L} =1$ scattering rate density with SM leptons, quarks and gauge bosons, $\gamma_{\mathrm{W},\Delta \mathrm{L}=1}^i$ is the rate of wash out processes caused by $\Delta \mathrm{L} =1$ scattering. We have checked that heavy neutrino mediated $\Delta \mathrm{L} =2$ scattering processes are negligible in our considered scenarios~\cite{Giudice:2003jh}.

Working in single flavor approximation with no spectator processes, the Boltzmann equations for the left-handed triplet decay scenarios are~\cite{Lavignac:2015gpa}
\begin{align}
sHz \frac{d\Sigma_\Delta}{dz} &= -\left( \frac{\Sigma_\Delta}{\Sigma_\Delta^{\rm eq}} -1 \right) \gamma_\mathrm{D} 
-2 \left[ \left(\frac{\Sigma_\Delta}{\Sigma_\Delta^{\rm eq}}\right)^2 -1 \right]\gamma_A;\\
sHz \frac{d\Delta_\Delta}{dz} &= -\left(  \frac{\Delta_\Delta}{\Sigma_\Delta^{\rm eq}}   -  B_l \frac{Y_\Delta}{Y_l^{\rm eq}} 
+  B_H \frac{Y_\Delta + 2 \Delta_\Delta}{Y_H^{\rm eq}}    \right) \gamma_\mathrm{D};\\
sHz \frac{Y_\Delta}{dz} &= -\left( \frac{\Sigma_\Delta}{\Sigma_\Delta^{\rm eq}} -1 \right) \gamma_\mathrm{D} \epsilon_\Delta
-2 B_l \left( \frac{dY_\Delta}{Y_l^{\rm eq}} -\frac{\Delta_{\Delta}}{\Sigma_{\Delta}^{\rm eq}}  \right) \gamma_\mathrm{D}
-2 \left( \frac{Y_\Delta}{Y_l^{\rm eq}} + \frac{Y_\Delta+ 2 \Delta_\Delta}{Y_H^{\rm eq}}  \right) \gamma_{lH},
\end{align}
where $\Sigma_\Delta \equiv (n_\Delta + n_{\bar{\Delta}})/s,~\Delta_\Delta \equiv (n_\Delta - n_{\bar{\Delta}})/s$, $\gamma_\mathrm{D}$ is the total decay rate density for the left-handed triplet scalar, $\gamma_A$ is the thermally averaged gauge scattering rate density for all ``$\Delta \mathrm{T} = 2$" processes ($\mathrm{T}$ referring to the left-handed triplet here), $\gamma_{lH}$ is the scattering rate density for $\Delta \mathrm{L} =2$ processes mediated by the left-handed triplet. The left-handed triplet has two decay channels and the branch ratios are
\begin{align}
B_l &\equiv \mathrm{BR}(\Delta \rightarrow \bar{l} \bar{l} )= \frac{m_\Delta^2\mathrm{Tr} \left( Y_\mathrm{T}^{} Y_\mathrm{T}^\dagger \right) }{m_\Delta^2 \mathrm{Tr} \left( Y_\mathrm{T}^{} Y_\mathrm{T}^\dagger \right)+|\mu|^2};\\
B_H &\equiv \mathrm{BR}(\Delta \rightarrow HH )=\frac{|\mu|^2 }{m_\Delta^2\mathrm{Tr} \left( Y_\mathrm{T}^{} Y_\mathrm{T}^\dagger \right)+|\mu|^2}.
\end{align}

\section{Results and discussions}\label{sec:result}

\subsection{Overview of the parameters and the numerical method}\label{subsec:overview}
\begin{table}[t!]
 \caption{\label{tab:glb} Neutrino oscillation parameters from Ref.~\cite{deSalas:2020pgw}.}\vspace{0.12cm}
 \centering
  \begin{tabular}{c|c|c}
   \toprule \hline 
    parameter & best fit & $1\sigma$ range\\ \hline 
    $\Delta m_{21}^2 $ [$10^{-5}$eV] & 7.50 & 7.30-7.72\\ 
    $\Delta m_{31}^2 $ [$10^{-3}$eV] (NO\footnotemark)& 2.56& 2.52-2.59\\
    $|\Delta m_{31}^2 |$ [$10^{-3}$eV] (IO)& 2.46 & 2.43-2.49\\
    $\sin^2 \theta_{12}/10^{-1}$ &3.18 & 3.02-3.34 \\
    $\sin^2 \theta_{23}/10^{-1}$  (NO) & 5.66 & 5.44 -5.82 \\
    $\sin^2 \theta_{23}/10^{-1}$  (IO) & 5.66  & 5.43-5.84 \\
    $\sin^2 \theta_{13}/10^{-2}$  (NO) & 2.225 & 2.147-2.28 \\
    $\sin^2 \theta_{13}/10^{-2}$  (IO) & 2.250  & 2.174-2.306 \\
    $\delta/\pi$  (NO) & 1.20 & 1.06-1.43 \\
    $\delta/\pi$ (IO)  & 1.54  & 1.41-1.67 \\
   \hline
  \bottomrule
\end{tabular}
\end{table}
\footnotetext{``NO (IO)" stands for normal (inverted) mass ordering of light neutrinos.}

It is helpful to discuss the model's relevant parameters before going into the details of the results. There are three sources of parameters:
\begin{enumerate}
\item In the light neutrino sector: the light neutrino masses and mixings. 
\item In the heavy neutrino sector: the heavy neutrino masses and mixings.
\item The rest: the triplet vevs, $v_\mathrm{L}$, $v_\mathrm{R}$, and the left-handed triplet mass $m_\Delta$.
\end{enumerate}

For the parameters in the  light neutrino sector, we adopt the current global fit values from Ref.~\cite{deSalas:2020pgw} and show them in Table~\ref{tab:glb} . This leaves us with only three unknowns: the lightest neutrino mass ($m$) and two Majorana phases ($\alpha_{21}, \alpha_{31}$). Our results are compatible with the low-energy observables in neutrino oscillation experiments with such a choice of input parameters.

For the heavy neutrino sector parameters, the lightest heavy neutrino mass is chosen by hand to set up the leptogenesis scale (in the heavy neutrino decay case) or to avoid complications (in the left-handed triplet decay case). We have checked that with all other parameters fixed, the heavy neutrino mass spectrum's influence on the numerical result is minor, as long as it is not degenerate and resonant regions are properly avoided. So we fix the heavy neutrino mass spectrum to be $m_{N_2} = 2 m_{N_1} ,~m_{N_3}  = 3 m_{N_1}$ (For a discussion of the heavy neutrino mass spectrum, see Appendix~\ref{apd:spectrum}). Since the mass spectrum is not very hierarchical, we consider all the three heavy neutrinos decay when they are lighter than the left-handed triplet. 
To minimize the free parameters, we take $V_\mathrm{R} = V_\mathrm{L}^*$ (when $\mathcal{C}$ is the LR symmetry) and $V_\mathrm{R} = V_\mathrm{L} $ (when $\mathcal{P}$ is the LR symmetry), which serves as a special case satisfying Eq.(\ref{eq:MD_C}) or Eq.(\ref{eq:MDH}). The coincidence of the light and heavy neutrino mixing works in type II seesaw limit and is not easily seen when type I seesaw contribution is included. This choice not only helps us to reveal the salient features of the model, which would otherwise be hindered by the complications in the right-handed neutrino mixing; but also leaves low-energy CP-violating phases the only source of  CP violation, which enables a direct connection between the low-energy CP violation and the BAU. The viability of the model with this choice well motivates further studies on a general basis.  We show the neutrino Dirac coupling in these cases in Table~\ref{tab:mD}. One can check that $M_\mathrm{D}^{} M_\mathrm{D}^\dagger$ is dependent on $V_\mathrm{L}$, which allows us to perform the unflavored leptogenesis analysis.

\begin{table}[t!]
 \caption{\label{tab:mD} Neutrino Dirac coupling matrix in our considered cases.}\vspace{0.12cm}
 \centering
  \begin{tabular}{c|c|c|c}
   \toprule \hline
  LR symmetry & RHN mixing\footnotemark & $M_\nu \simeq M_\nu^\mathrm{I}$ & $M_\nu \simeq M_\nu^\mathrm{II}$\\ \hline
   $\mathcal{P}$ & $V_\mathrm{R} = V_\mathrm{L} $& $M_\mathrm{D}= i V_\mathrm{L} V_\mathrm{L} \sqrt{\mathrm{D}_\nu \mathrm{D}_N} V_\mathrm{L}^\dagger$ & $ \displaystyle M_\mathrm{D}= V_\mathrm{L} V_\mathrm{L} \mathrm{D}_N\sqrt{\frac{v_\mathrm{L}}{v_\mathrm{R}}-\mathrm{D}_\nu^{} \mathrm{D}_N^{-1}} V_\mathrm{L}^\dagger$ \\
   $\mathcal{C}$ & $V_\mathrm{R} = V_\mathrm{L}^* $ & $M_\mathrm{D}= i V_\mathrm{L}^*V_\mathrm{L}^* \sqrt{\mathrm{D}_\nu \mathrm{D}_N} V_\mathrm{L}^\dagger$ & $ \displaystyle M_\mathrm{D}= V_\mathrm{L}^* V_\mathrm{L}^* \mathrm{D}_N\sqrt{\frac{v_\mathrm{L}}{v_\mathrm{R}}-\mathrm{D}_\nu^{} \mathrm{D}_N^{-1}} V_\mathrm{L}^\dagger$\\
   \hline
  \bottomrule
\end{tabular}
\end{table}
\footnotetext{Right-handed neutrino mixing matrix.}
For the rest of the model parameters, the left-handed triplet mass $m_\Delta$ is fixed in the same spirit as the lightest heavy neutrino mass. The ratio $v_\mathrm{L}/v_\mathrm{R}$ is severely constrained once we choose the way of light neutrino mass generation (type I or type II domination). See a detailed description in Appendix~\ref{sec:appdx1}.
To avoid the damping effects from the $W_\mathrm{R}$ related processes in leptogenesis, $W_\mathrm{R}$ mass has to be ~\cite{Ma:1998sq,Frere:2008ct}
\begin{align}
M_{W_\mathrm{R}} > (2\times 10^5 \mathrm{GeV} ) \left( \frac{M_{N_1}}{10^2 \mathrm{GeV}} \right)^{3/4}.
\end{align}
This in turn sets a lower limit on $v_\mathrm{R}$ as $v_\mathrm{R} >10^{13}$ GeV.

Note that the parameters we fix also change the predictions of BAU. Their effects are not our primary concern for the following reasons. First, for the heavy neutrino spectrum, as is mentioned, as long as it is not (nearly-) degenerate, it affects the final BAU slightly and thus is neglected; Second, for the masses that set the leptogenesis scale, they do alter the prediction greatly. We mainly focus on the low-energy CP phases' contribution to the final BAU, which can be justified like this: the investigation here can be viewed as a fixed-heavy mass slide of the whole parameter space, and any positive result on this slide can motivate further investigation on other slides.

To sum up, due to the structure given by MLRSM, we do not have much freedom with the parameters relevant to leptogenesis in this framework (except for the heavy neutrino mixing). We assume the heavy and light neutrino mixing coincidence (up to conjugation) to reduce free parameters further and enable low-energy CP violation the only CP source for leptogenesis. It is really interesting to see what we can get for the resulting baryon asymmetry. 

As it is known that solving the Boltzmann equations is time-consuming, we have no intention to investigate the whole viable parameter space, which is formidable in this framework with all these scenarios. Instead, we choose to do it in a ``scattered plot" way. We consider the following two scenarios for numerical analysis
\begin{itemize}
\item Scenario A: fix $\alpha_{21}=\alpha_{31}=0$, vary $m \in [0.0001,0.03] $ eV and $\delta \in  [0, 2 \pi]$. In this scenario, the Dirac CP phase is the sole source of CP violation.
\item Scenario B: fix $m=0.01$ eV and $\delta=\delta_\mathrm{bf}$ ($\delta_\mathrm{bf}$ is the best fit value for $\delta$ in the considered mass ordering), vary $\alpha_{21}\in [0, 2 \pi]$ and $\alpha_{31}\in [0, 2 \pi]$. In this scenario the Majorana phases can contribute.
\end{itemize}
The rest neutrino sector parameters are set to their best fit values.

Different light neutrino mass generation mechanisms and lepton asymmetry generation mechanisms give four cases as mentioned in Sec.~\ref{sec:lep}.
Taking into consideration also the discrete LR symmetry and neutrino mass ordering, we have $16$ cases as shown in Table~\ref{tab:class}.  We name the models as follows: the first letter (P, C) represents the LR symmetry as $\mathcal{P}$ or $\mathcal{C}$; the second number (1,2,3,4) denotes the case among the four defined in Sec.~\ref{sec:lep}; the last two letters (NO, IO) represent the light neutrino mass ordering.

\begin{table}[t!]
 \caption{\label{tab:class} Classification of the $16$ cases.}\vspace{0.12cm}
 \centering
  \begin{tabular}{c|c|c|c}
   \toprule \hline
  lepton asymmetry & neutrino mass & $\mathcal{P}$ as the LR symmetry & $\mathcal{C}$ as the LR symmetry\\ \hline
  \multirow{2}{*}{N decay} & type I & P1NO, P1IO & C1NO, C1IO\\ 
  &type II &P2NO, P2IO & C2NO, C2IO\\
   \hline
   \multirow{2}{*}{$\Delta$ decay} & type I & P3NO, P3IO & C3NO, C3IO\\
  &type II &P4NO, P4IO & C4NO, C4IO\\
   \hline
  \bottomrule
\end{tabular}
\end{table}

\subsection{Type I mass domination, heavy neutrino decay }

\begin{figure}
\centering
\includegraphics[width=\textwidth]{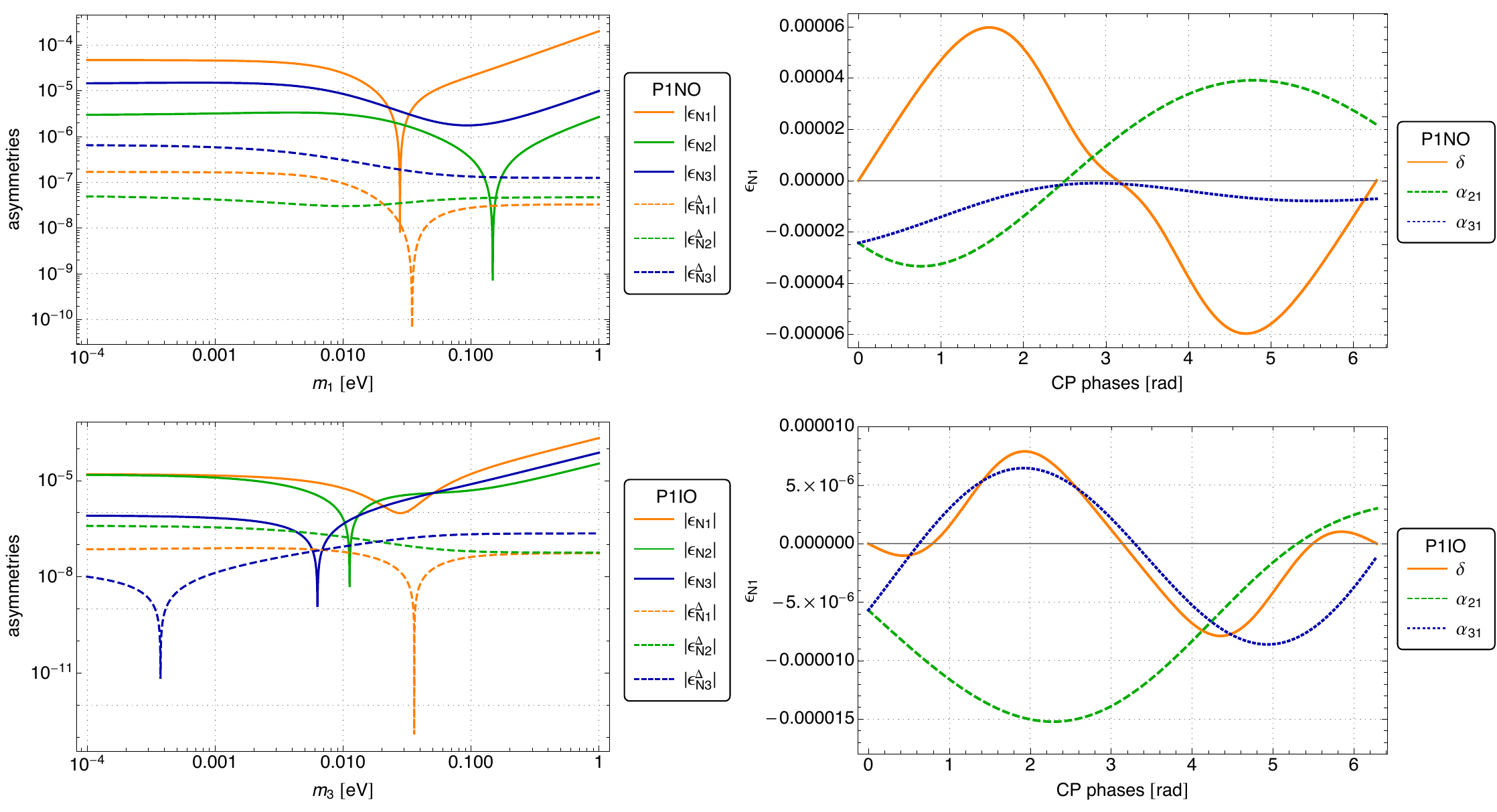}
\includegraphics[width=\textwidth]{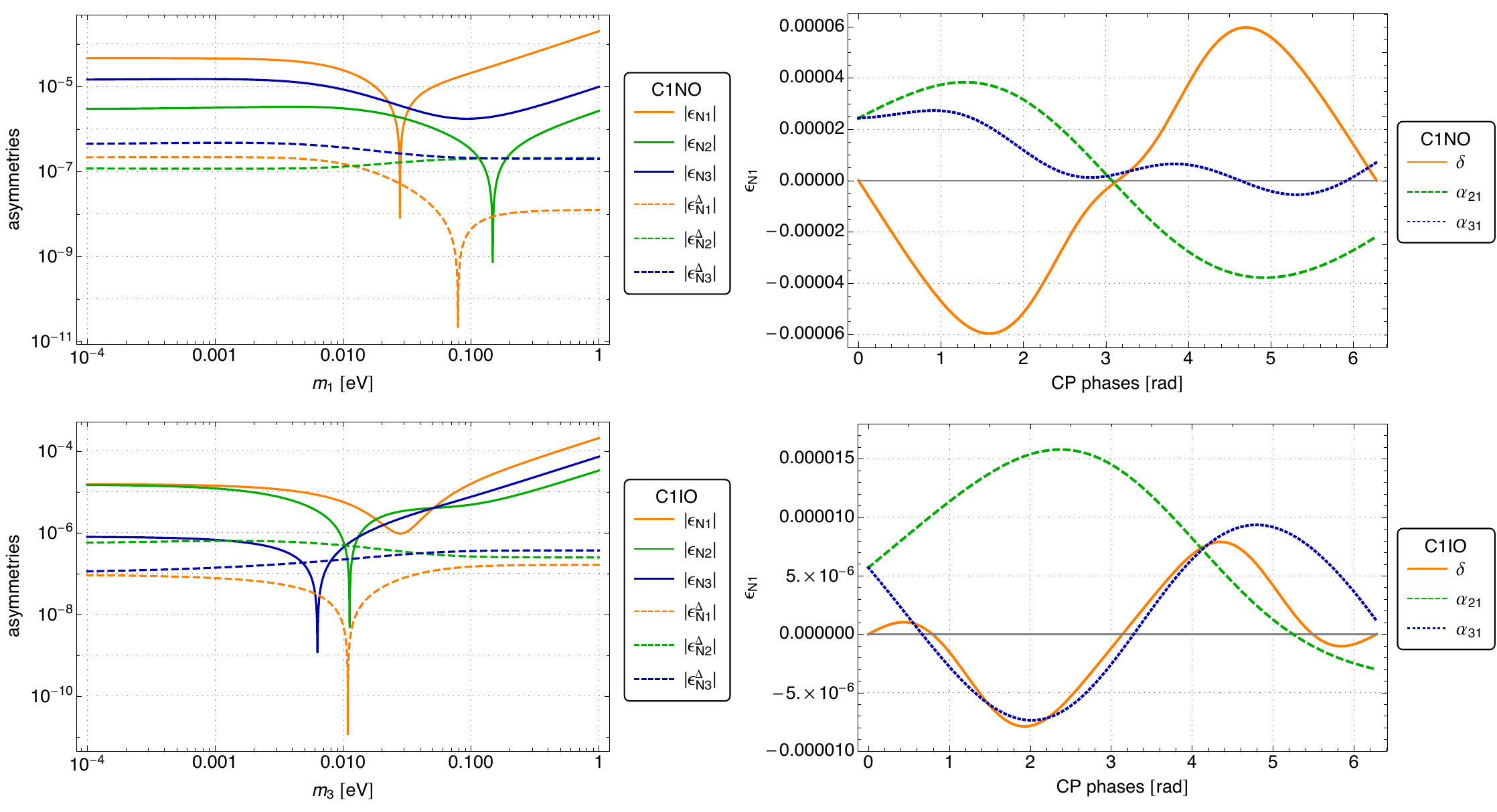}
\caption{The CP asymmetries generated by heavy neutrino decay as a function of the lightest light neutrino mass (left column) and the CP-violating phases (right column) in cases P1NO, P1IO, C1NO, and C1IO. Left column: the solid lines are asymmetries generated by interference with the first two diagrams in Fig.~\ref{fig:Ndecay}, while the dotted lines represent the contribution with a left-handed triplet in the loop (the third diagram in Fig.~\ref{fig:Ndecay}). The Dirac phase is fixed to its best fit value, while the Majorana phases are set to zeros. Right column: the solid line represents variation with the Dirac CP phase by setting the Majorana phases to zeros; the dashed and dotted lines denote variations with the Majorana phases with the Dirac CP phase setting to its best fit value. The lightest neutrino mass is set to $0.01$ eV. The rest parameters are fixed either at their best fit values or otherwise stated in the text.
}
\label{fig:ePlot1}
\end{figure}

\begin{figure}
\centering
\includegraphics[width=0.6\textwidth]{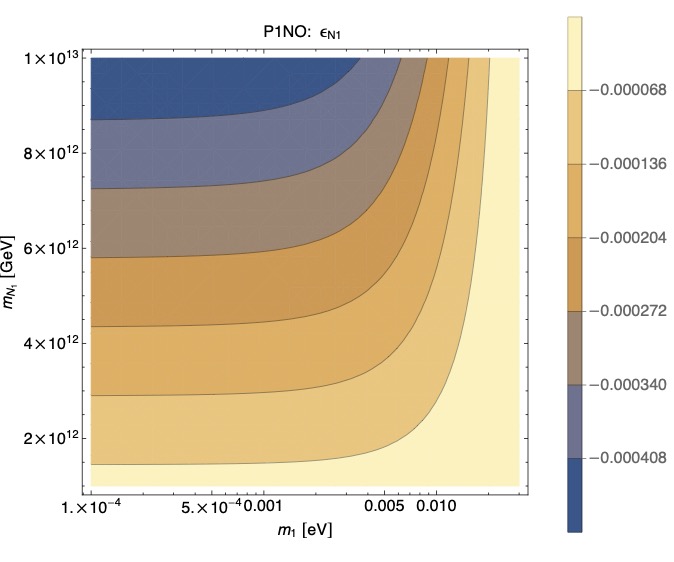}
\caption{The CP asymmetry $\epsilon_{N_1}$ as a function of the lightest light neutrino mass $m_1$ and the lightest heavy neutrino mass $m_{N_1}$ in the  case P1NO. The oscillation parameters are fixed at their best fit values and the Majorana phases are set to zeros.}
\label{fig:mmPlotp1no}
\end{figure}

\begin{figure}
\centering
\includegraphics[width=0.9\textwidth]{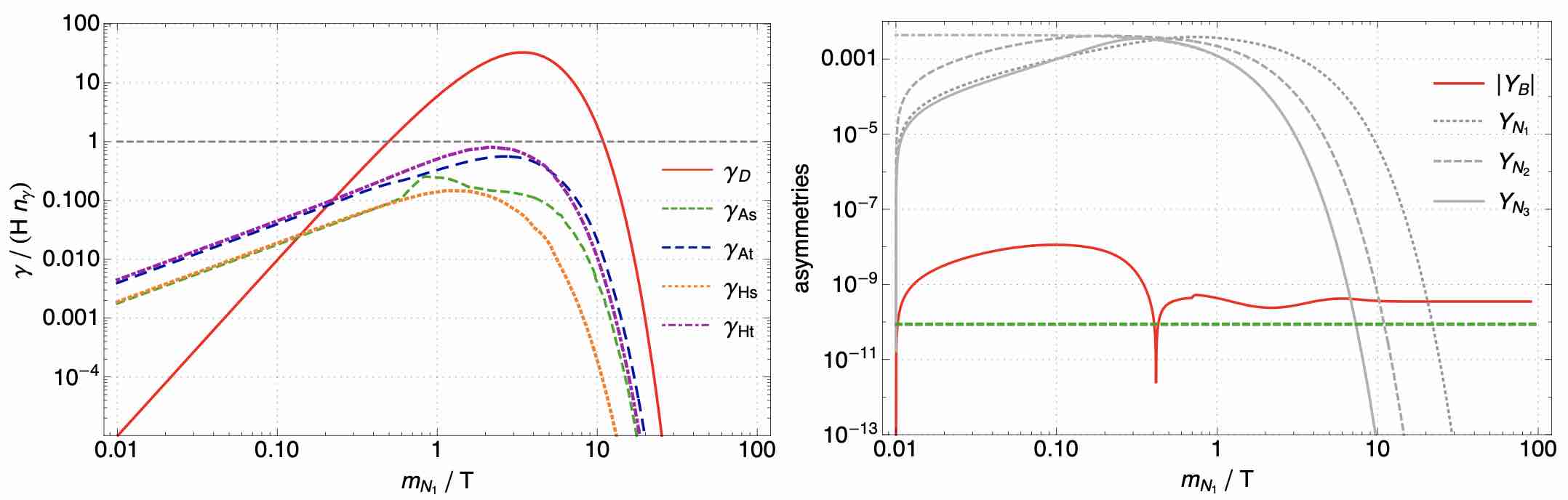}
\caption{Left: The interaction densities as a function of $z=m_{N_1}/T$ in the case P1NO. Right: The baryon asymmetry and heavy neutrino comoving number densities as a function of $z=m_{N_1}/T$ in the case P1NO. The horizontal green dashed lines are the $3\sigma$ region of observed baryon asymmetry $Y_\mathrm{B}=(8.72\pm 0.08)\times 10^{-11}$~\cite{Aghanim:2018eyx}. The oscillation parameters are fixed at their best fit values, and the Majorana phases are set to zeros. $m_1$ is fixed to $0.01$ eV.}
\label{fig:ratePlotp1no}
\end{figure}

\begin{figure}
\centering
\includegraphics[height=.45\textheight]{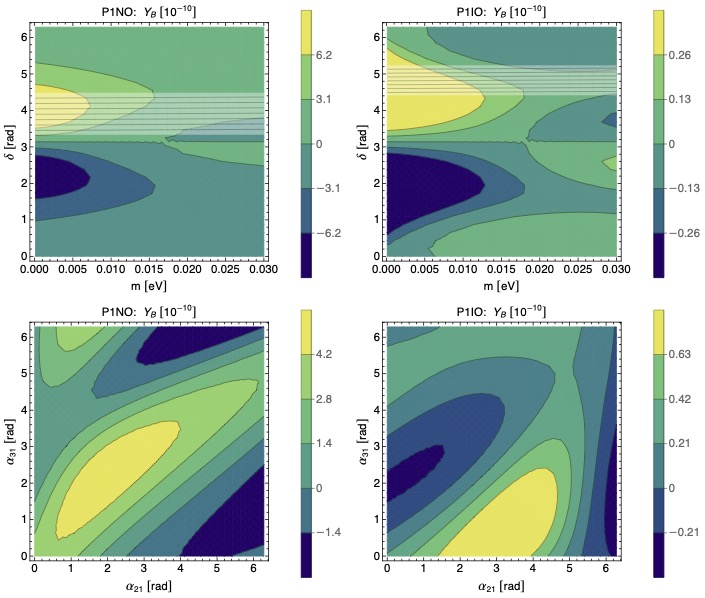}
\includegraphics[height=.45\textheight]{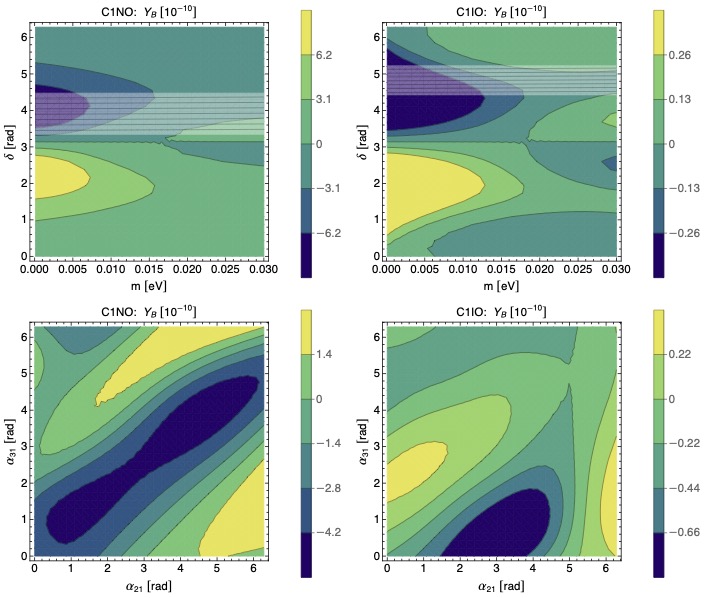}
\caption{The resulting $Y_\mathrm{B}$ for two variables in cases P1NO, P1IO, C1NO, and C1IO. In the first and the third row the Dirac phase $\delta$ is the only CP-violating source (The Majorana phases are fixed to zeros). The horizontal white mesh band is the current global fit $1\sigma$ region of $\delta$. The second and the fourth row show $Y_\mathrm{B}$ with variations of the Majorana phases, where $\delta$ is fixed to its best fit value.
}
\label{fig:result1}
\end{figure}

In the P1NO and the P1IO cases, we set
\begin{align}
m_{N_1}=10^{12}~ \mathrm{GeV},~m_\Delta = 10^{14}~\mathrm{GeV},~\displaystyle \frac{v_\mathrm{L}}{v_\mathrm{R}} = 10^{-25}.
\end{align}
%which gives rise to $\overbar{M_\nu^\mathrm{I}}/\overbar{M_\nu^\mathrm{II}} \simeq 10^2$. 
Although $M_\mathrm{D}$ is different in the C1NO and the C1IO cases, when we fix the third category parameters at these value, we are still in type I mass dominated region. As a result, these parameters are left untouched in the C1NO and the C1IO cases.

We plot the CP asymmetries generated by heavy neutrino decay as a function of the interested low-energy parameters in Fig.~\ref{fig:ePlot1}. Generally, one expects a CP asymmetry larger than $\mathcal{O}(10^{-6})$ to be able to reproduce the magnitude of baryon asymmetry. It is the case as shown in this plot. We see from the left column that in general the asymmetry generated by the first two diagrams in Fig.~\ref{fig:Ndecay} are $1.5$ - $2$ orders larger than that generated by the third diagram. It is a result of the type I mass dominating the light neutrino mass. Under this constraints, the triplet lepton coupling are typically at $\mathcal{O}(10^{-4})$, while the neutrino Dirac coupling (which now makes main contribution to light neutrino mass) is at $\mathcal{O}(10^{-2} )$.

Since we mainly focus on the low-energy parameters, we fix the heavy neutrino masses to be $m_{N_2} = 2 m_{N_1} ,~m_{N_3}  = 3 m_{N_1} $in our numerical discussion. It should be addressed that the heavy neutrino masses, especially the lightest one, is also crucial to the final baryon asymmetry. It affects not only the CP asymmetry but also the efficiency of the ``conversion" to the final baryon asymmetry (as a variable for various interaction densities). We plot the CP asymmetry as a function of the lightest light neutrino mass and heavy neutrino mass in Fig.~\ref{fig:mmPlotp1no} as an illustration of its effect. Note that its effect on the final baryon asymmetry cannot be compensated by the light neutrino mass variation. The most important feature that comes with a varying $m_{N_1}$ is entering the flavored regime. In that regime, the charged lepton Yukawas enter into equilibrium and become distinguishable. Including the flavor effect may allow a sufficient baryon asymmetry generation with $m_{N_1}$ several orders lower. It is beyond the scope of the current work and deserves a separate study.

Given our input, within the whole considered range of the lightest neutrino mass, $m \in [10^{-4},0.003]$ eV, the tree level total decay width of $N_1$ is larger than the expansion rate of the Universe, i.e., $\Gamma_{\mathrm{D}_{N_1}}/ H(T=m_{N_1}) \sim \mathcal{O}(10)$. As a result, the inverse decay and scatterings mediated by the same Dirac coupling are efficient, and we are in a strong washout regime. We show a typical interaction density and comoving number density evolution in Fig.~\ref{fig:ratePlotp1no}.

We perform a numerical study on the baryon asymmetry's dependence on the low-energy parameters, the method is described in the previous section, and the results are shown in Fig.~\ref{fig:result1}. From these plots, we see that the observed baryon asymmetry can be accommodated in various cases. Current global fit $1\sigma$ region of $\delta$ ``choose" the correct sign of $Y_\mathrm{B}$ for most regions of $m$ when $\mathcal{P}$ is the LR symmetry in both orderings. It means the Dirac CP phase alone can offer the right amount of baryon asymmetry $Y_\mathrm{B}$: not only the magnitude but also the correct sign. The resulting baryon asymmetry spans a larger region when the light neutrino mass ordering is normal than inverted. The baryon asymmetry generally has opposite signs for cases with $\mathcal{P}$ as the LR symmetry and its counterpart with $\mathcal{C}$ as the LR symmetry. The Majorana phases in the inverted ordering open up viable parameter space as the additional CP-violating sources. However, it is not the case for normal ordering.

\subsection{Type II mass domination, heavy neutrino decay }

\begin{figure}[t!]
\centering
\includegraphics[width=\textwidth]{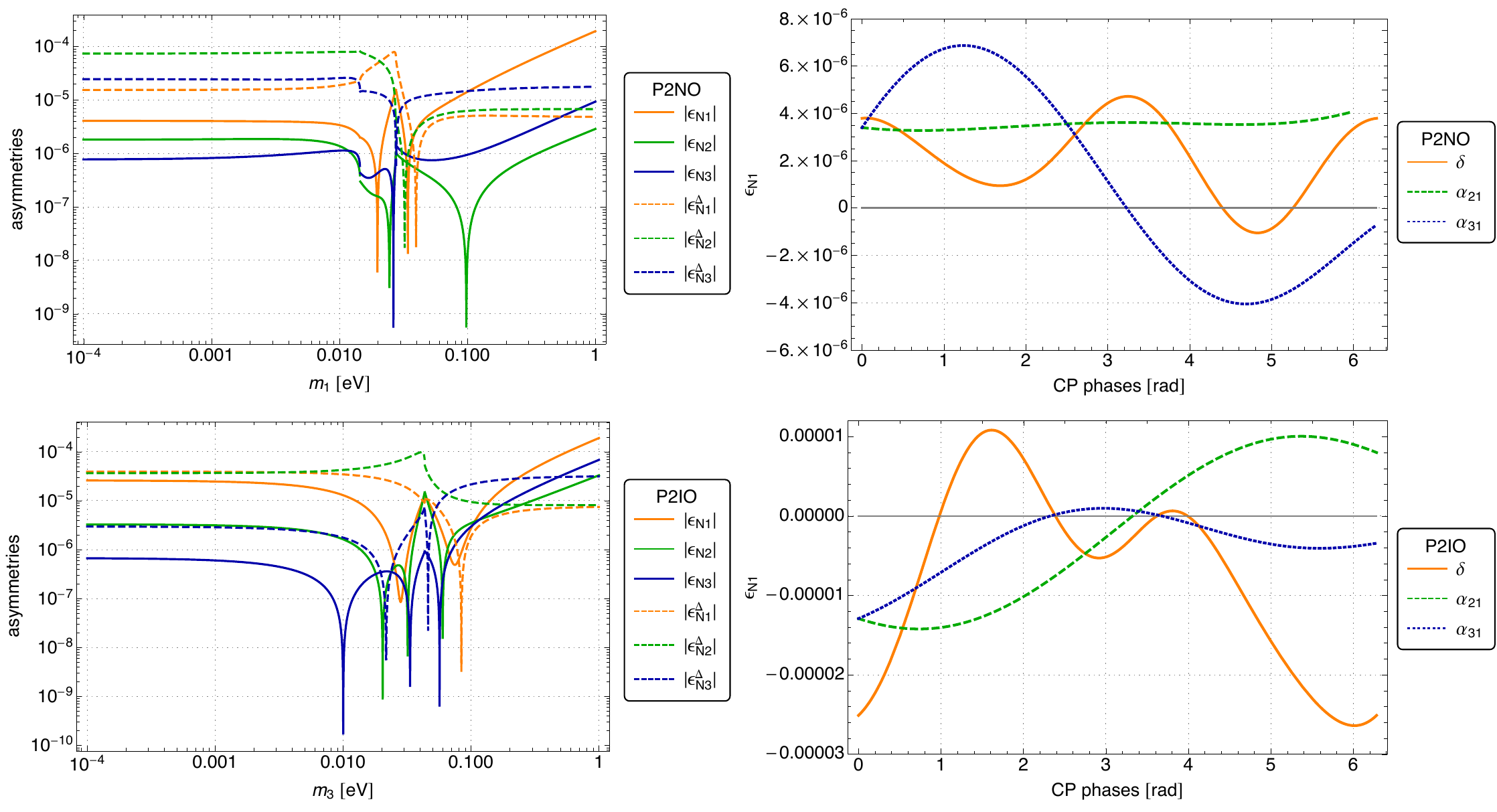}
\includegraphics[width=\textwidth]{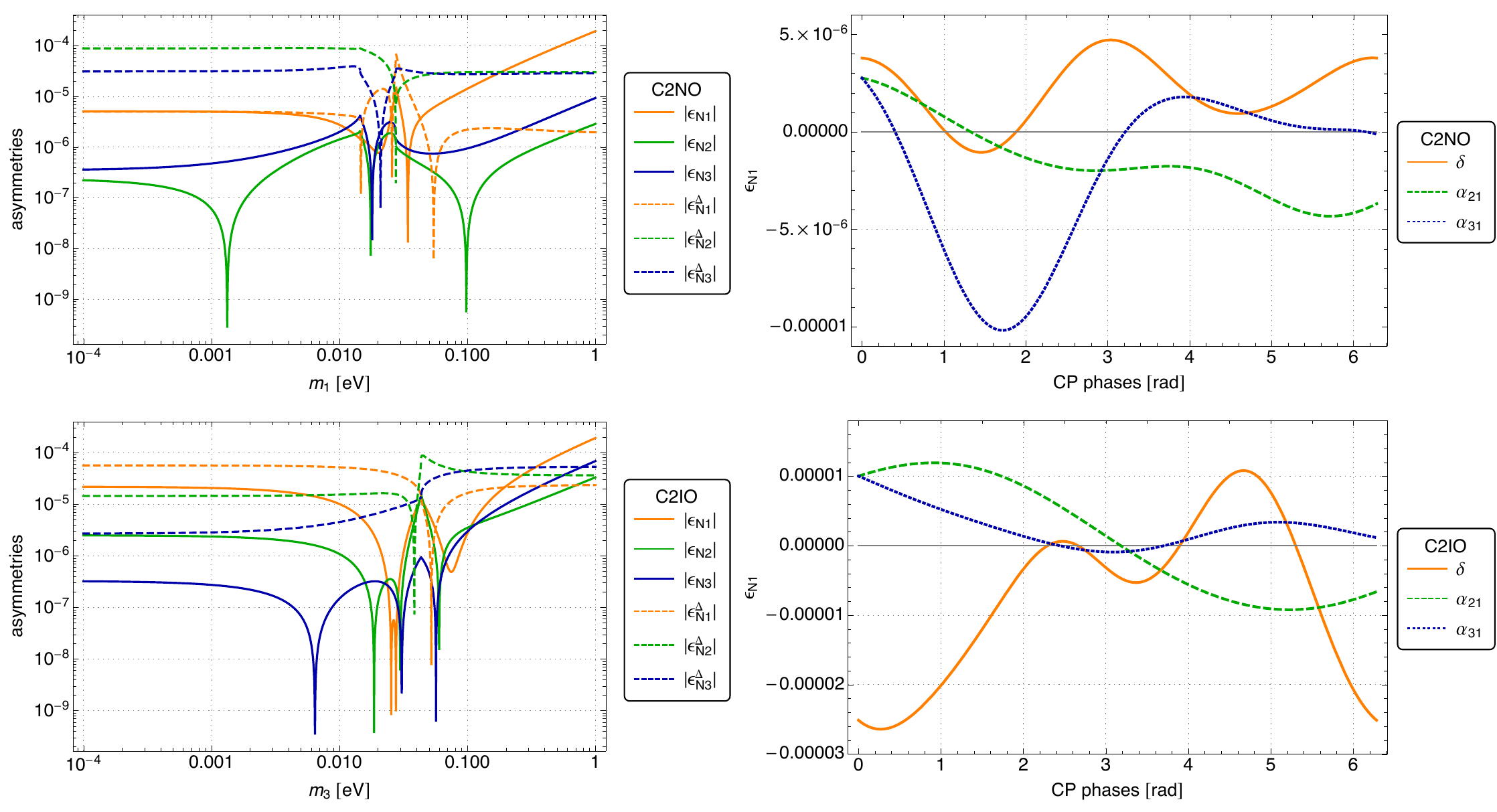}
\caption{The CP asymmetries generated by heavy neutrino decay as a function of the lightest light neutrino mass (left column) and the CP-violating phases (right column) in cases P2NO, P2IO, C2NO, and C2IO. Left column: the solid lines are asymmetries generated by interference with the first two diagrams in Fig.~\ref{fig:Ndecay}, while the dotted lines represent the contribution with a left-handed triplet in the loop (the third diagram in Fig.~\ref{fig:Ndecay}). The Dirac phase is fixed to its best fit value, while the Majorana phases are set to zeros. Right column: the solid line represents variation with the Dirac CP phase by setting the Majorana phases to zeros; the dashed and dotted lines denote variations with the Majorana phases with the Dirac CP phase setting to its best fit value. The lightest neutrino mass is set to $0.01$ eV. The rest parameters are fixed either at their best fit values or otherwise stated in the text.
}
\label{fig:ePlot2}
\end{figure}

\begin{figure}
\centering
\includegraphics[height=.45\textheight]{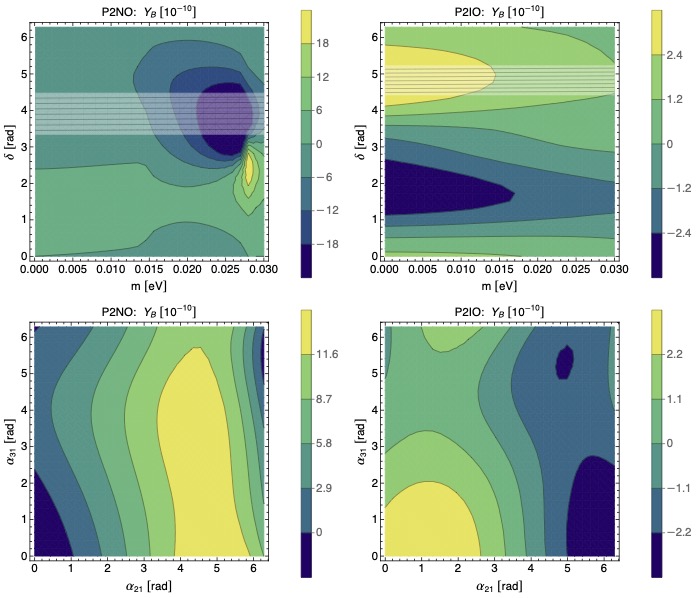}
\includegraphics[height=.45\textheight]{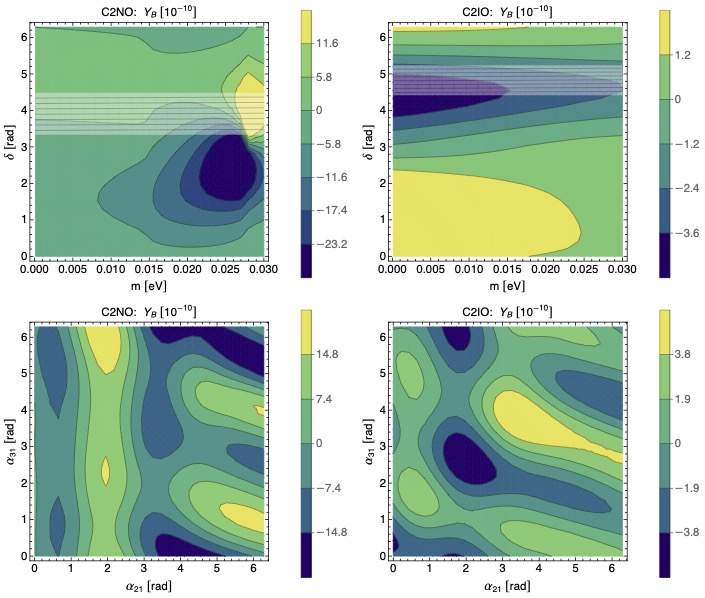}
\caption{The resulting $Y_\mathrm{B}$ for two variables in cases P2NO, P2IO, C2NO, and C2IO. In the first and the third row the Dirac phase $\delta$ is the only CP-violating source (The Majorana phases are fixed to zeros). The horizontal white mesh band is the current global fit $1\sigma$ region of $\delta$. The second and the fourth row show $Y_\mathrm{B}$ with variations of the Majorana phases, where $\delta$ is fixed to its best fit value.
}
\label{fig:result2}
\end{figure}

In this subsection, we set
\begin{align}
m_{N_1}=10^{12}~ \mathrm{GeV},~m_\Delta = 4\times10^{13}~\mathrm{GeV},~\displaystyle \frac{v_\mathrm{L}}{v_\mathrm{R}} =1.45\times 10^{-23}.
\end{align}
The ratio $v_\mathrm{L}/v_\mathrm{R}$ is set by requiring a type II mass dominance in mixed type I$+$II mechanism (See Appendix~\ref{sec:appdx1} for more details). With this ratio of $v_\mathrm{L}/v_\mathrm{R}$, by requiring that $\mu/v_\mathrm{R} < 1$, the left-handed triplet mass has an upper limit $\displaystyle m_\Delta < v v_\mathrm{R}/v_\mathrm{L} = 4.56 \times 10^{13}$ GeV. 

We show the CP asymmetries generated by heavy neutrino decay as a function of the lightest light neutrino mass and the CP-violating phases in cases P2NO, P2IO, C2NO, and C2IO in Fig.~\ref{fig:ePlot2}. As the type II mass dominates the light neutrino mass, we see now $\epsilon_{N_1}^\Delta$ or $\epsilon_{N_2}^\Delta$  is the largest in most of the interested lightest light neutrino mass range, i.e., $m<0.03$ eV as set by the Planck 2018 result. This fact can be illustrated when we consider the hierarchical heavy neutrino mass spectrum~\cite{Hambye:2003ka},
\begin{align}
\epsilon_{N_1} &= \displaystyle \frac{3}{16\pi} \frac{m_{N_1}}{v^2} \frac{\sum_{i,l} \mathrm{Im} [ (Y_N)_{1i} (Y_N)_{1l} (M_\nu^{\mathrm{I}*})_{il}] }{ \sum_i | (Y_N)_{1i}|^2},\\
\epsilon_{N_1}^\Delta &= \displaystyle -\frac{1}{8\pi} \frac{m_{N_1}}{v^2} \frac{\sum_{i,l} \mathrm{Im} [ (Y_N)_{1i} (Y_N)_{1l} (M_\nu^{\mathrm{II}*})_{il}] }{ \sum_i | (Y_N)_{1i}|^2}. 
\end{align}
If entities in $M_\nu^\mathrm{II}$ are large enough, it is possible to have $\epsilon_{N_1}^\Delta > \epsilon_{N_1}^{} $ as we see in Fig.~\ref{fig:ePlot2}.

The Dirac coupling is different from previous cases, but for our chosen parameters, the ratio $\Gamma_{\mathrm{D}_{N_1}}/ H(T=m_{N_1})$ is still at $ \mathcal{O}(10)$, although smaller than that in previous cases. We do not show the interaction densities here as they look similar.

The numerical results for the baryon asymmetry are shown in Fig.~\ref{fig:result2}, from which we find all cases can accommodate an observed baryon asymmetry and many features similar to previous cases. The resulting $Y_\mathrm{B}$ in normal ordering spans a larger range than inverted ordering. The baryon asymmetry generally has opposite signs for cases with $\mathcal{P}$ as the LR symmetry and its counterpart with $\mathcal{C}$ as the LR symmetry. 
There are also some differences. The inverted ordering in scenario A of case P2IO and the normal ordering case in scenario A of C2NO is now favored as they coincidence the right sign with $1\sigma$ region of the Dirac phase.

\subsection{Type I mass domination, left-handed triplet scalar decay }

\begin{figure}
\centering
\includegraphics[width=0.6\textwidth]{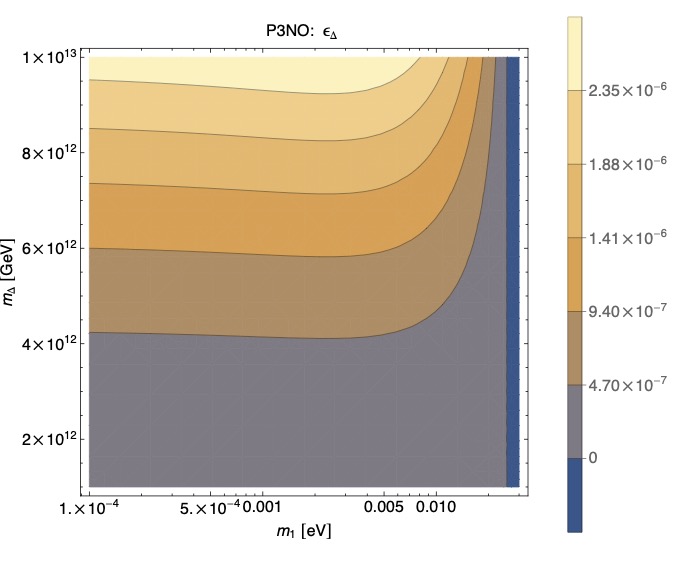}
\caption{The CP asymmetry $\epsilon_\Delta$ as a function of the lightest light neutrino mass $m_1$ and the left-handed triplet mass $m_\Delta$ in the case P3NO. The oscillation parameters are fixed at their best fit values and the Majorana phases are set to zeros.}
\label{fig:mmPlotp3no}
\end{figure}

\begin{figure}
\centering
\includegraphics[width=0.9\textwidth]{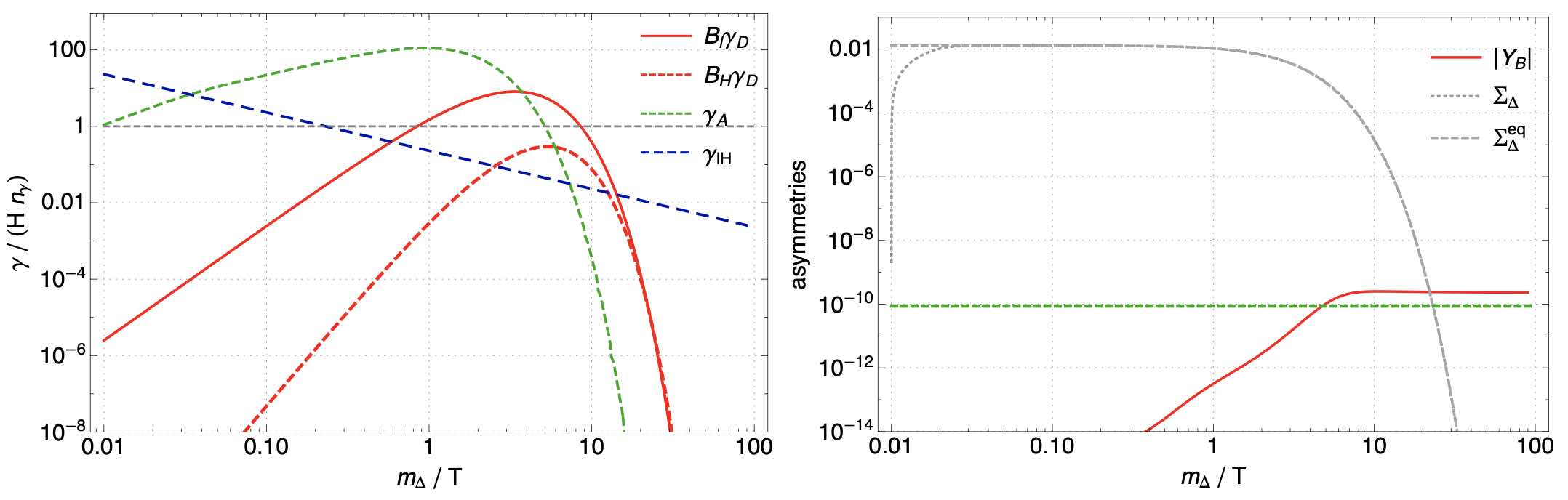}
\caption{Left: The interaction rate densities as a function of $z=m_\Delta/T$ in the case P3NO. Right: The baryon asymmetry and the sum of comoving left-handed triplets number densities as a function of $z=m_\Delta/T$ in the case P3NO. The horizontal green dashed lines are the $3\sigma$ region of observed baryon asymmetry $Y_\mathrm{B}=(8.72\pm 0.08)\times 10^{-11}$~\cite{Aghanim:2018eyx}. The oscillation parameters are fixed at their best fit values, and the Majorana phases are set to zeros. $m_1$ is fixed to $0.01$ eV.}
\label{fig:ratePlotp3no}
\end{figure}

\begin{figure}
\centering
\includegraphics[width=\textwidth]{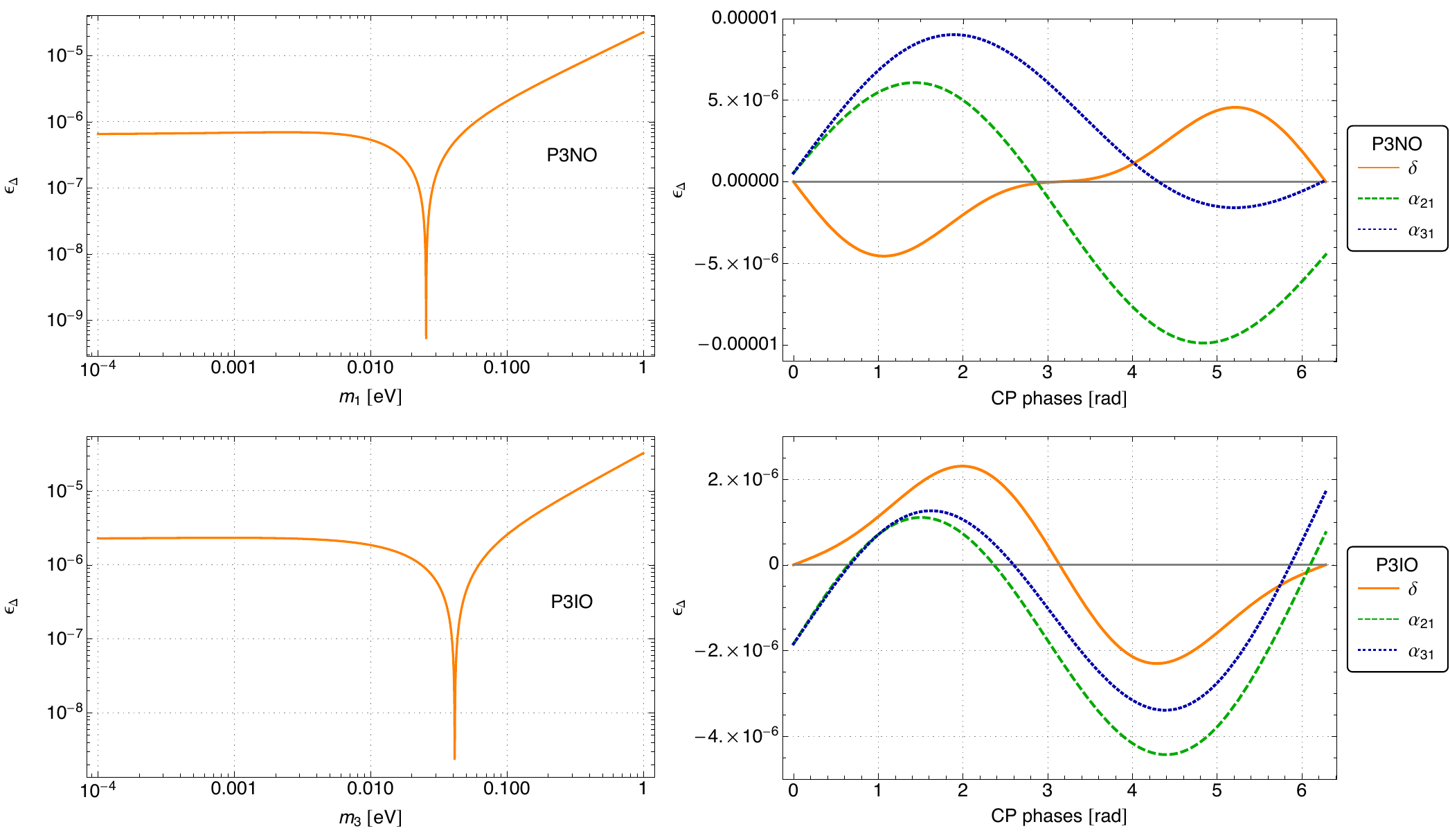}
\includegraphics[width=\textwidth]{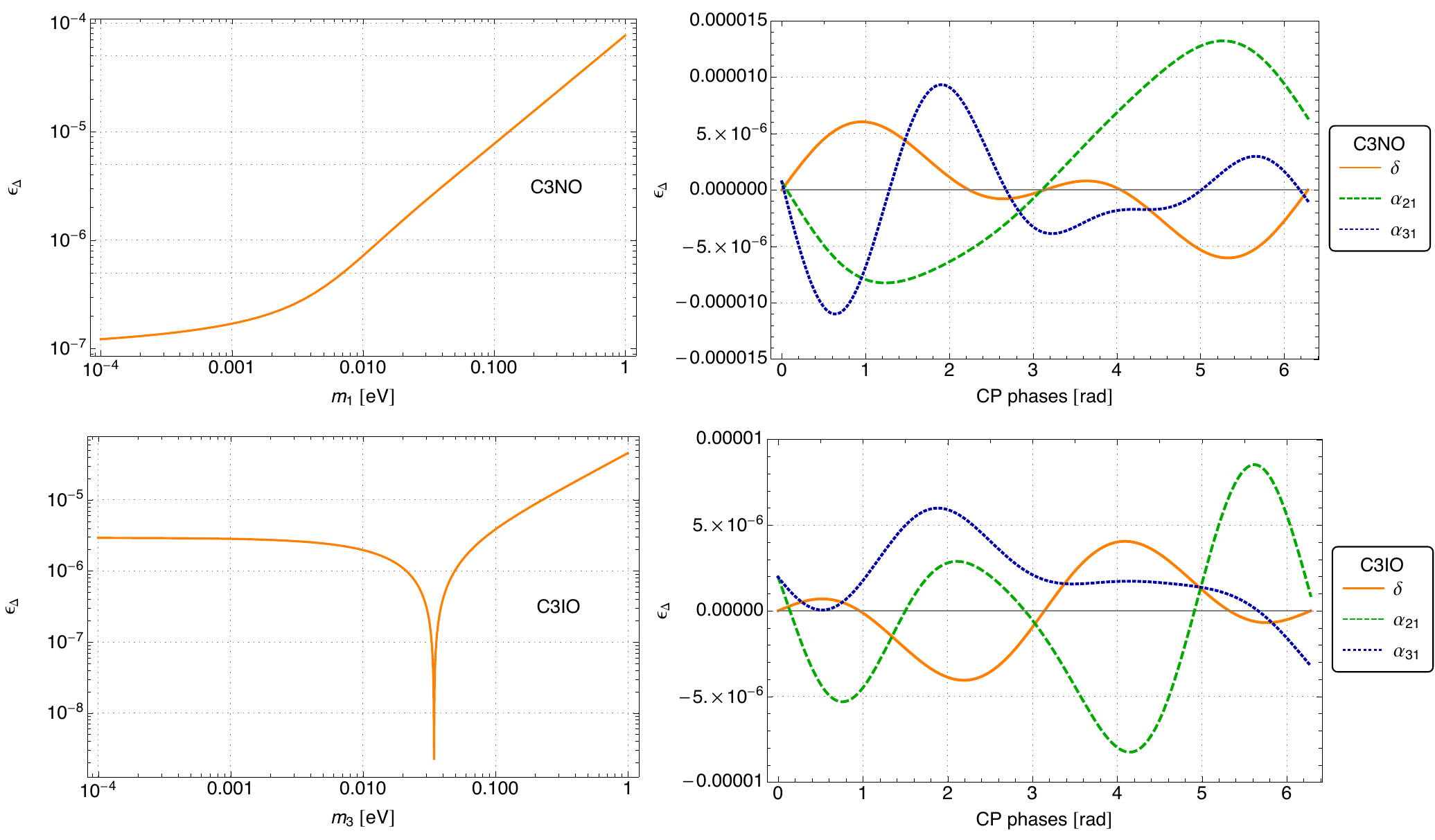}
\caption{The CP asymmetry generated by left-handed triplet decay as a function of the lightest neutrino mass (left column) and the CP-violating phases (right column) in cases P3NO, P3IO, C3NO, and C3IO. Left column: the Dirac phase is fixed to its best fit value, while the Majorana phases are set to zeros. Right column: the solid line represents variation with the Dirac CP phase by setting the Majorana phases to zeros; the dashed and dotted lines denote variations with the Majorana phases with the Dirac CP phase setting to its best fit value. The lightest neutrino mass is set to $0.01$ eV. The rest parameters are fixed either at their best fit values or otherwise stated in the text.
}
\label{fig:ePlot3}
\end{figure}

\begin{figure}
\centering
\includegraphics[height=.45\textheight]{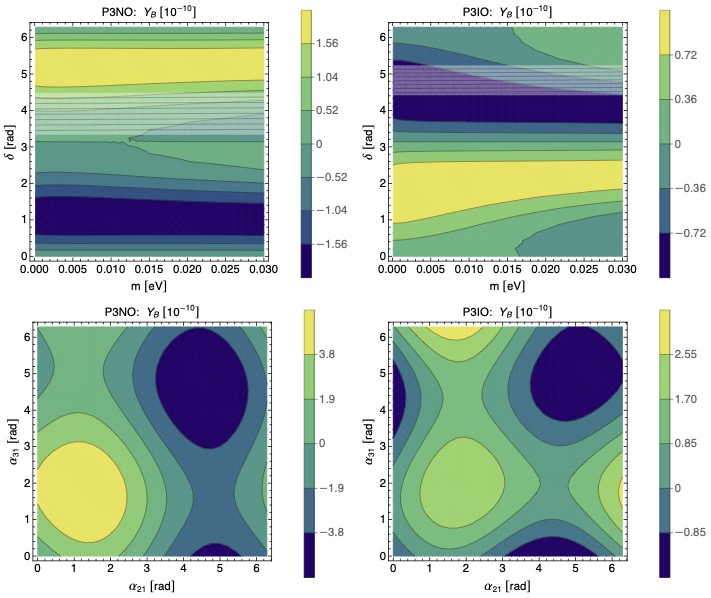}
\includegraphics[height=.45\textheight]{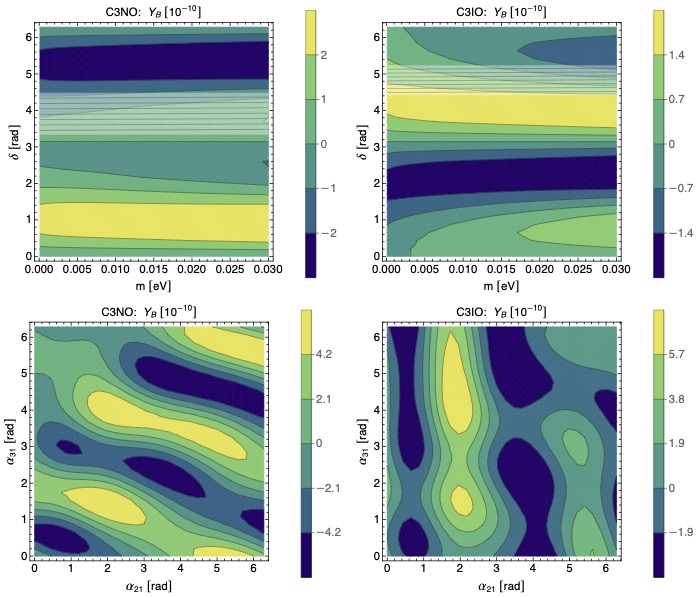}
\caption{The resulting $Y_\mathrm{B}$ for two variables in cases P3NO, P3IO, C3NO, and C3IO. In the first and the third row the Dirac phase $\delta$ is the only CP-violating source (The Majorana phases are fixed to zeros). The horizontal white mesh band is the current global fit $1\sigma$ region of $\delta$. The second and the fourth row show $Y_\mathrm{B}$ with variations of the Majorana phases, where $\delta$ is fixed to its best fit value.
}
\label{fig:result3}
\end{figure}

In this subsection, we take the following input 
\begin{align}
m_{N_1}= 10^{14} \mathrm{GeV},~m_\Delta = 5\times 10^{12} \mathrm{GeV},~v_\mathrm{L}=10^{-11} \mathrm{GeV},~v_\mathrm{R}=10^{16} \mathrm{GeV}.
\end{align}
This $v_\mathrm{L}/v_\mathrm{R}$ gives $\overbar{M}_\nu^\mathrm{I}/\overbar{M}_\nu^\mathrm{II} \simeq 203$ for $m=0.01$ eV in cases P3NO, P3IO, and  $\overbar{M}_\nu^\mathrm{I}/\overbar{M}_\nu^\mathrm{II} \simeq 207$ in cases C3NO, C3IO.

Although we fix the left-handed triplet mass in our numeric study, it plays an important role in generating the final baryon asymmetry. We show the CP asymmetry as a function of the lightest neutrino mass and the left-handed triplet mass in Fig.~\ref{fig:mmPlotp3no}. A larger CP asymmetry favors a smaller lightest neutrino mass but a larger left-handed triplet mass slightly.

As the left-handed triplet has two decay channels, it is sufficient to have one channel satisfy the out-of-equilibrium condition. From the interaction rate density plot in Fig.~\ref{fig:ratePlotp3no}, we see that the left-handed triplet decays to Higgs is out-of-equilibrium. It also shows both the $\Delta \mathrm{L} =2$ and gauge scatterings are fast at first, but as time evolves, decay starts dominating, and a net lepton asymmetry or equivalently a baryon asymmetry grows and survives. 
We have the left-handed triplet decay branching ratios $B_l \simeq 0.998,~ B_H \simeq 0.002$ in these cases, which leads to a maximal efficiency when converting to final baryon asymmetry~\cite{Hambye:2005tk}. Indeed, we see from Fig.~\ref{fig:ratePlotp3no} that one of the two decay rates (decay to leptons) is faster than the gauge scattering rate, which makes gauge scattering ineffective; the decay to two Higgs is slower than the expansion rate, making the lepton asymmetry survive.

We show the CP asymmetry generated by left-handed triplet decay as a function of the lightest neutrino mass (left column) and the CP-violating phases (right column) in cases P3NO, P3IO, C3NO, and C3IO in Fig.~\ref{fig:ePlot3}. The CP asymmetry is continuous in these parameters. The variations with CP phases show multiple periods when $\mathcal{C}$ is the LR symmetry and correspond to multiple maxima and minima in the last row of Fig.~\ref{fig:result3}. The numerical results of baryon asymmetry are shown in Fig.~\ref{fig:result3}. 

\subsection{Type II mass domination, left-handed triplet scalar decay }

\begin{figure}
\centering
\includegraphics[width=\textwidth]{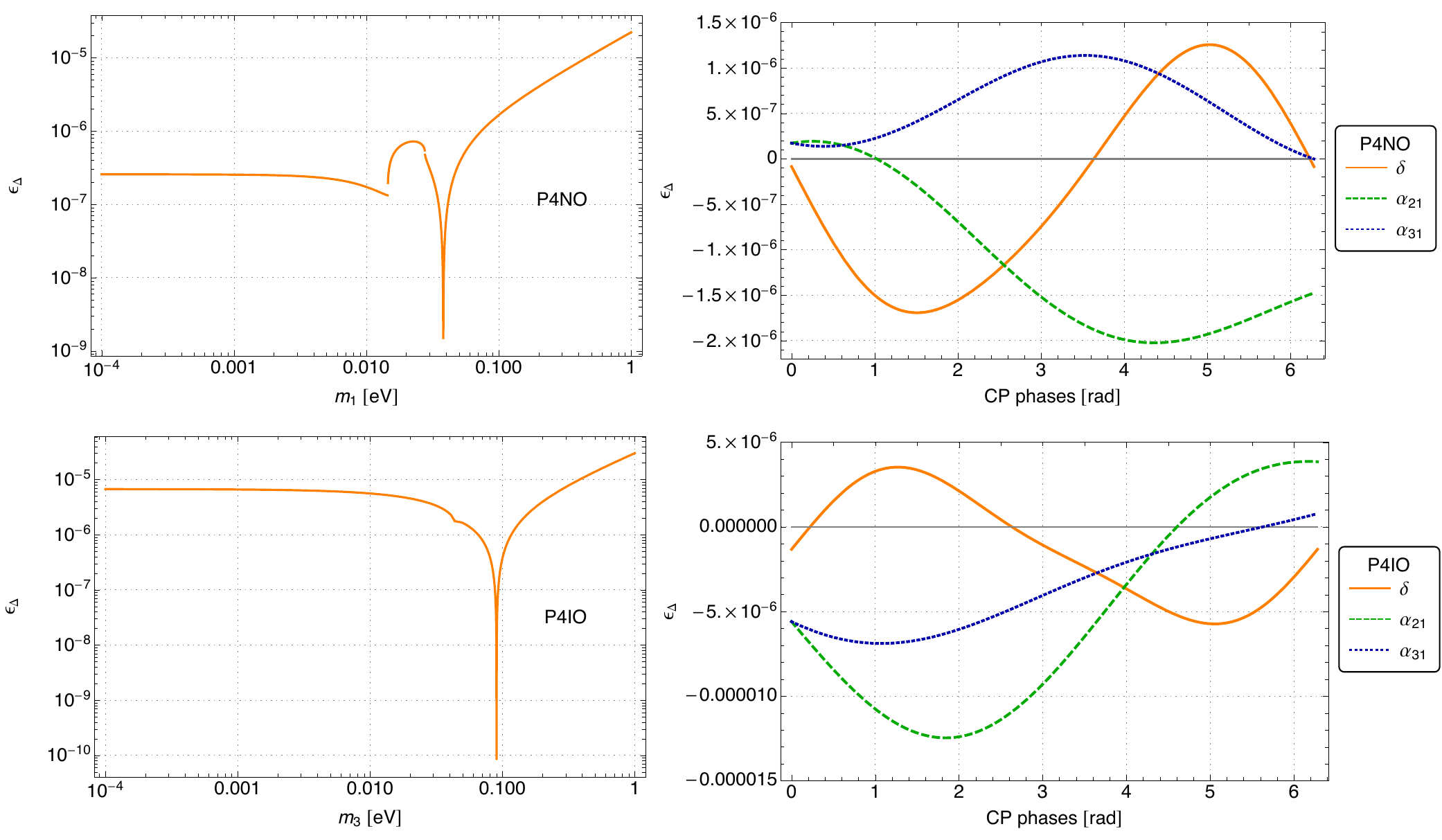}
\includegraphics[width=\textwidth]{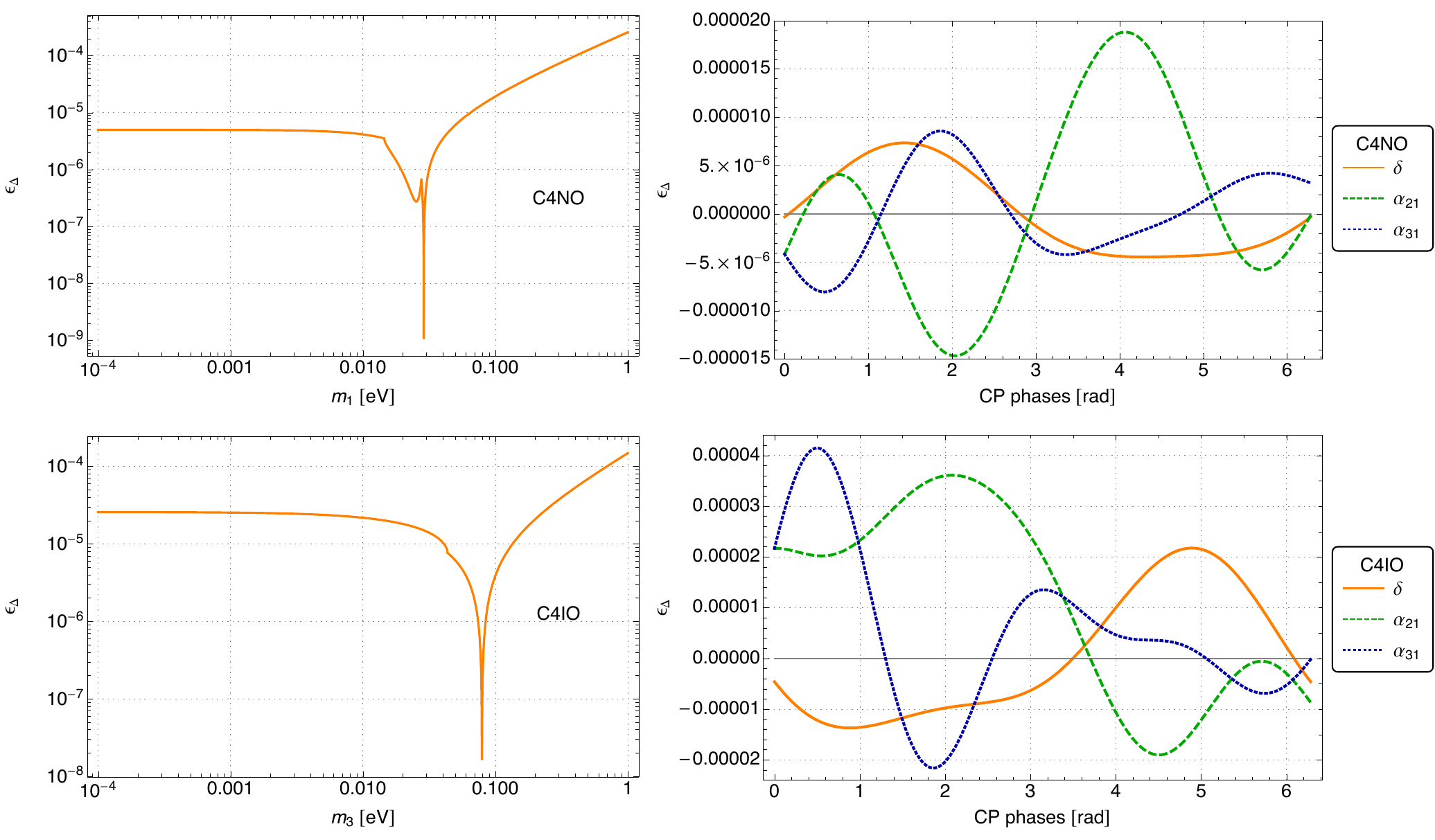}
\caption{The CP asymmetries generated by left-handed triplet decay as a function of the lightest neutrino mass (left column) and the CP-violating phases (right column) in cases P4NO, P4IO, C4NO, and C4IO. Left column: the Dirac phase is fixed to its best fit value, while the Majorana phases are set to zeros. Right column: the solid line represents variation with the Dirac CP phase by setting the Majorana phases to zeros; the dashed and dotted lines denote variations with the Majorana phases with the Dirac CP phase setting to its best fit value. The lightest neutrino mass is set to $0.01$ eV. The rest parameters are fixed either at their best fit values or otherwise stated in the text.
}
\label{fig:ePlot4}
\end{figure}

\begin{figure}
\centering
\includegraphics[height=.45\textheight]{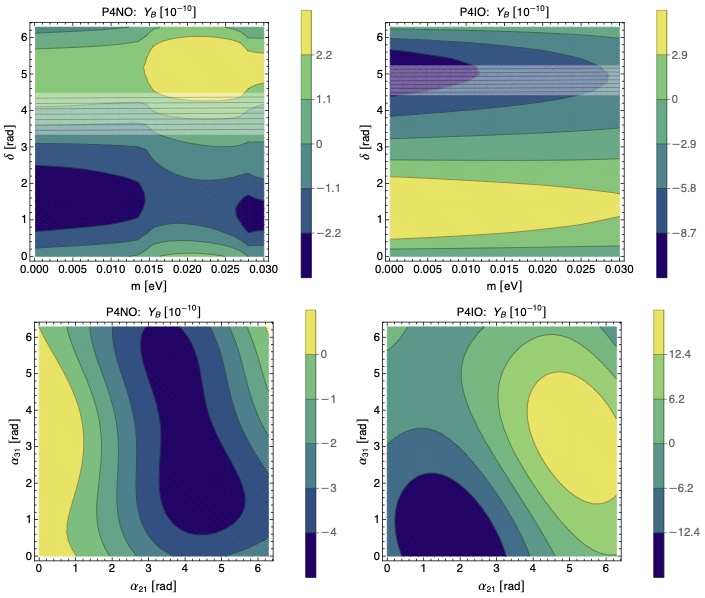}
\includegraphics[height=.45\textheight]{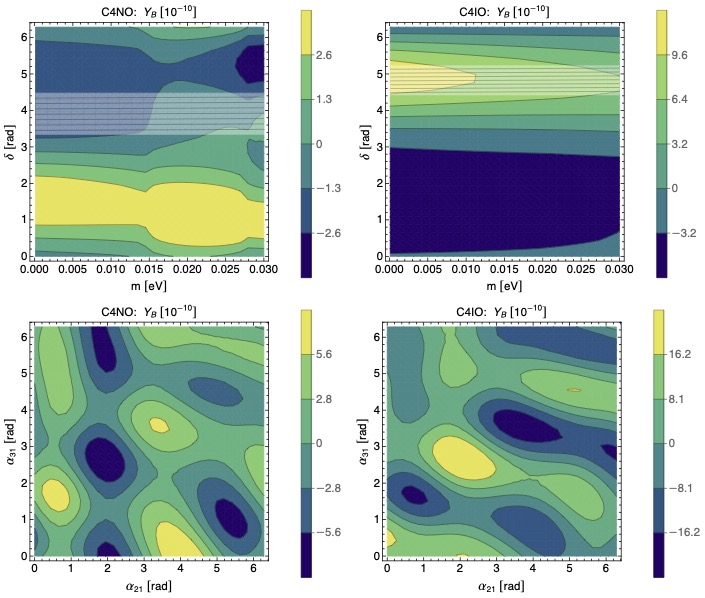}
\caption{The resulting $Y_\mathrm{B}$ for two variables in cases P4NO, P4IO, C4NO, and C4IO. In the first and the third row the Dirac phase $\delta$ is the only CP-violating source (The Majorana phases are fixed to zeros). The horizontal white mesh band is the current global fit $1\sigma$ region of $\delta$. The second and the fourth row show $Y_\mathrm{B}$ with variations of the Majorana phases, where $\delta$ is fixed to its best fit value.
}
\label{fig:result4}
\end{figure}

\begin{figure}
\centering
\includegraphics[width=0.9\textwidth]{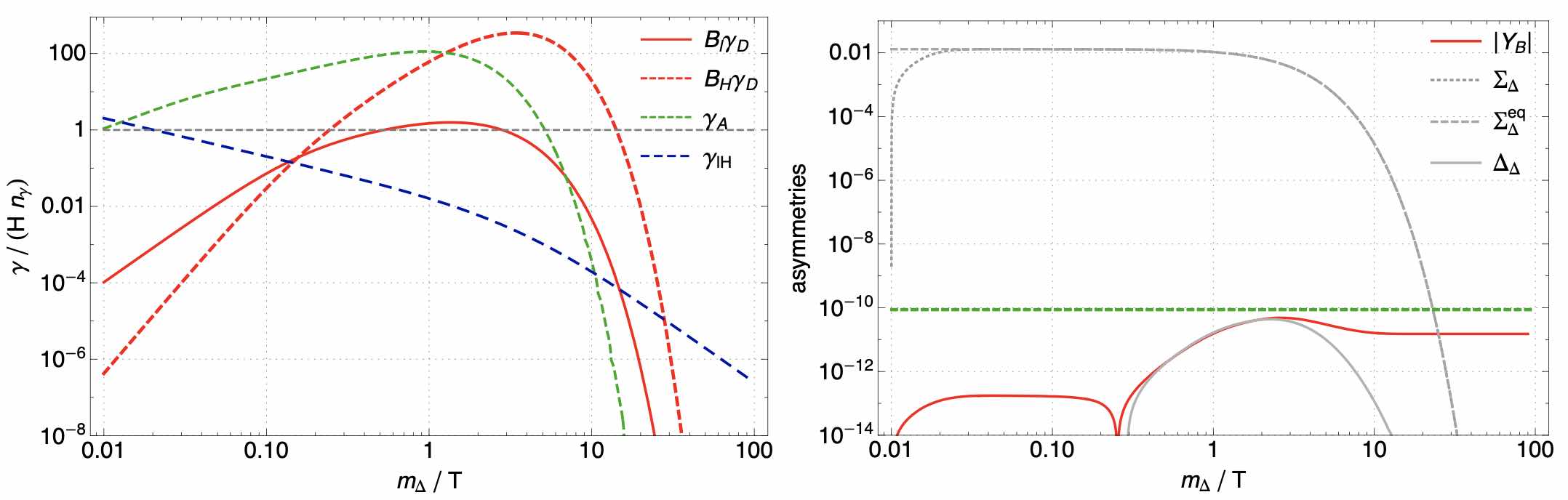}
\caption{Left: The interaction rate densities as a function of $z=m_\Delta/T$ in the case P4NO. Right: The baryon asymmetry and the sum of comoving left-handed triplets number densities as a function of $z=m_\Delta/T$ in the case P4NO. The horizontal green dashed lines are the $3\sigma$ region of observed baryon asymmetry $Y_\mathrm{B}=(8.72\pm 0.08)\times 10^{-11}$~\cite{Aghanim:2018eyx}. The oscillation parameters are fixed at their best fit values, and the Majorana phases are set to zeros. $m_1$ is fixed to $0.01$ eV.}
\label{fig:ratePlotp4no}
\end{figure}

In this subsection, we set
\begin{align}
m_{N_1}=10^{14}~ \mathrm{GeV},~m_\Delta = 5\times10^{12}~\mathrm{GeV},~v_\mathrm{L} =1.45\times 10^{-9}~\mathrm{GeV},~v_\mathrm{R} = 10^{16}~\mathrm{GeV}.
\end{align}
This $v_\mathrm{L}/v_\mathrm{R}$ is again (as P2NO, P2IO cases) set by $\overbar{M_\nu}/\overbar{M_\nu^\mathrm{II}}= 1.01$ (Note that $m_{N_1}$ is two-orders larger than that in the P2NO case).

We have the left-handed triplet decay branching ratios $B_l \simeq 0.024,~ B_H \simeq 0.976$ in these cases, i.e., the left-handed triplet mainly decay into Higgs. Such branch ratios also lead to a maximal efficiency when converting to final baryon asymmetry~\cite{Hambye:2005tk}. As in cases discussed in the previous subsection, we see from Fig.~\ref{fig:ratePlotp4no} that one of the two decay rates (decay to Higgs) is faster than the gauge scattering rate, which makes gauge scattering ineffective; the decay to two leptons is now slower than expansion rate, making the lepton asymmetry survive. %$\Gamma_\Delta/ H(T=m_\Delta) \simeq 83$. It is decaying to leptons goes out of equilibrium.

We show the CP asymmetry generated by left-handed triplet decay as a function of the lightest neutrino mass (left column) and the CP-violating phases (right column) in cases P4NO, P4IO, C4NO, and C4IO in Fig.~\ref{fig:ePlot4} and the numerical results in Fig.~\ref{fig:result4}. We see that the observed baryon asymmetry can be generated in all these cases. P4NO and C4IO are favored as they have the right sign chosen by $1\sigma$ region of the Dirac CP phase. In variation with the lightest neutrino mass, the CP asymmetries generated in cases C4NO and C4IO are almost one order larger than those of C3NO, C3IO. In variations with CP phases, the CP asymmetries again show multiple maxima and minima, which can also be observed in the last row of Fig.~\ref{fig:result4}.

\section{Interplay among low-energy CP violation, leptogenesis and neutrinoless double beta decay}\label{sec:interplay}

\begin{figure}[h!]
\centering
\includegraphics[width=0.9\textwidth]{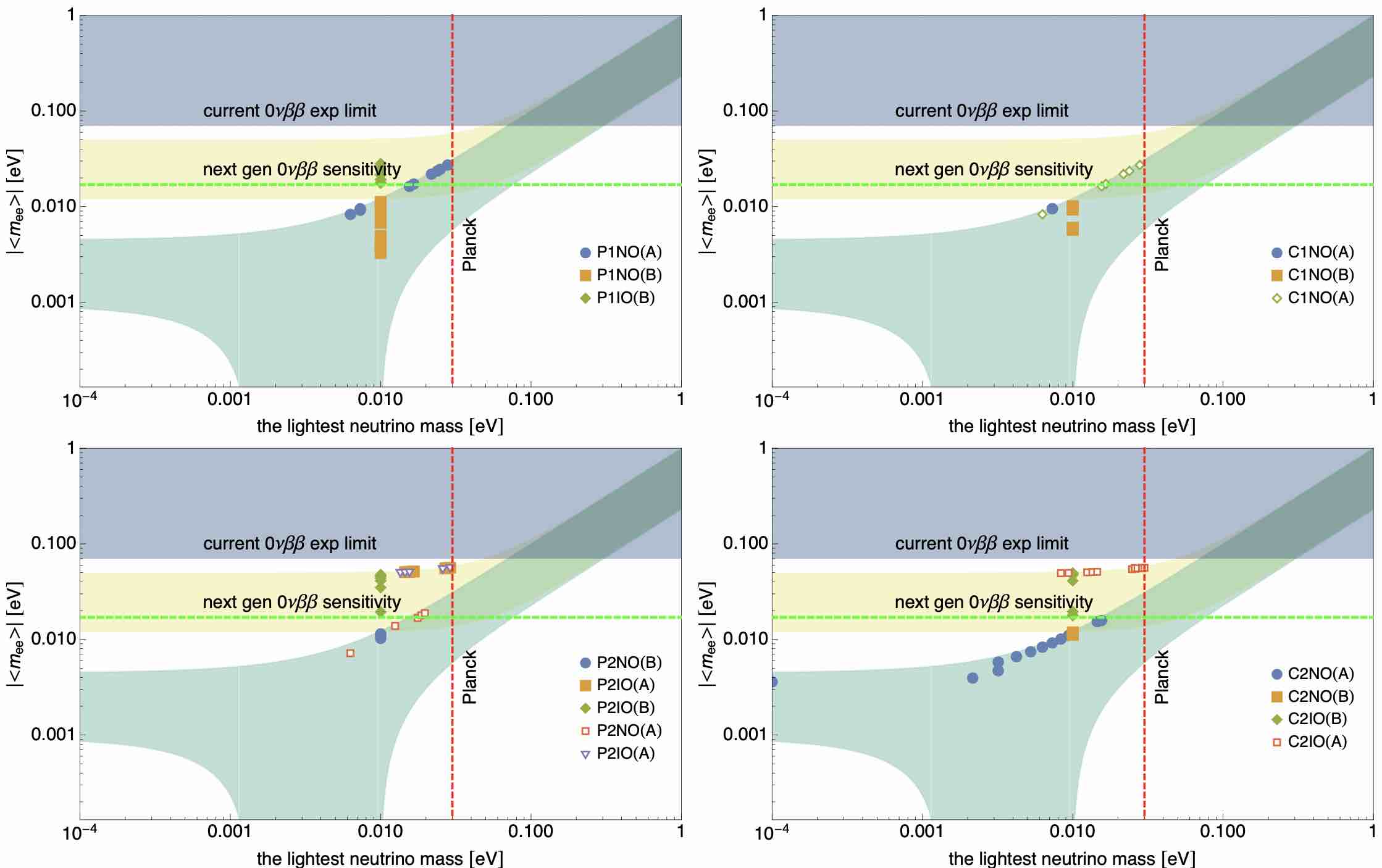}
\includegraphics[width=0.9\textwidth]{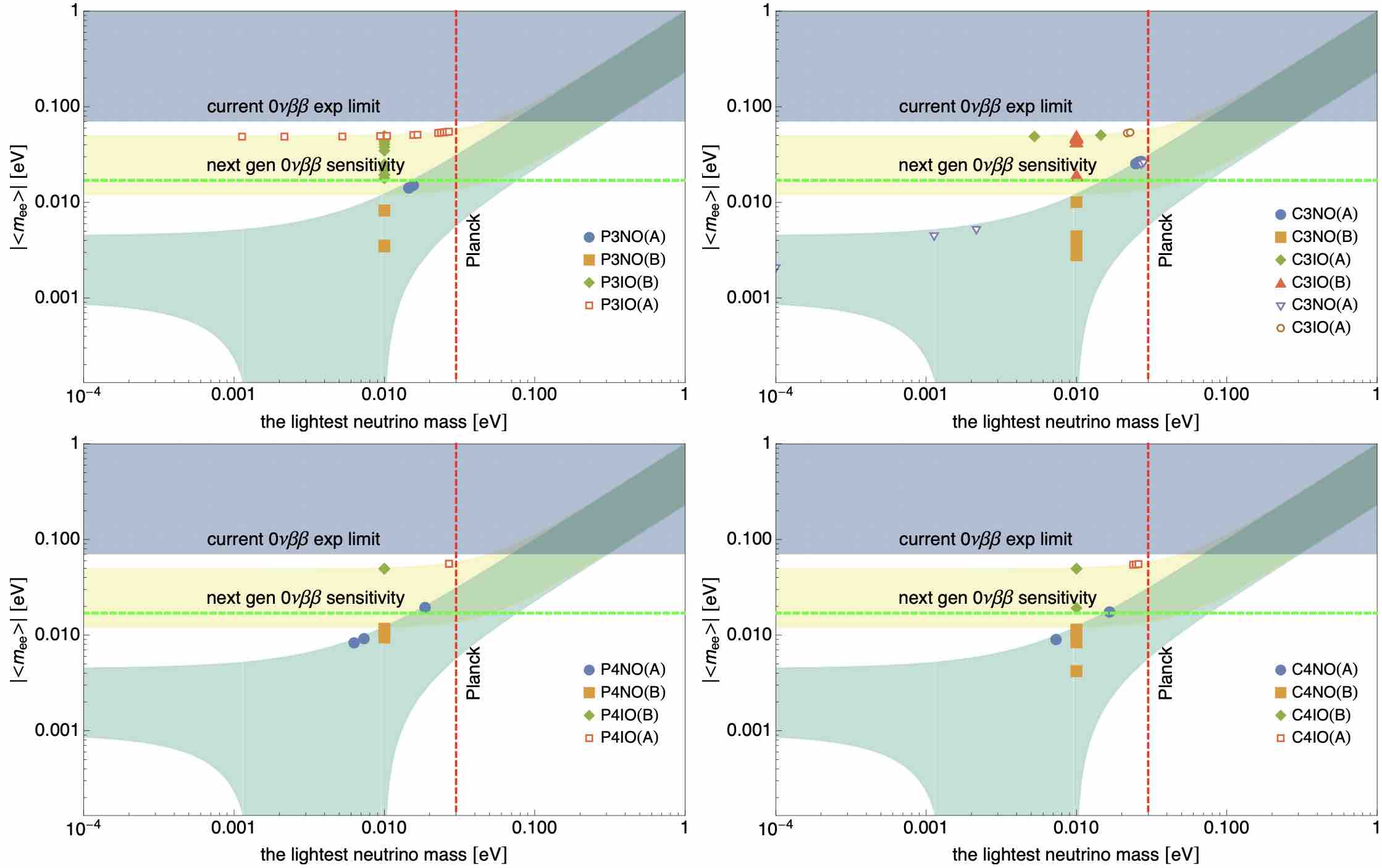}
\caption{The effective Majorana mass with dependence on the lightest neutrino mass. The shown points are taken from our numerical scan by the criteria that they generate a baryon asymmetry within the $3\sigma$ region of the observed value. We take the combined limit set by current neutrinoless double beta decay experiments from Ref.~\cite{GERDA2019}. For next-generation experiment sensitivity denoted by the green horizontal dashed line, we use nEXO~\cite{Albert:2017hjq} as an example. We also show the Planck limit as the vertical red dashed line. The filled points are within the $3\sigma$ ranges of current oscillation parameters. Note that the scenario B points are all in these ranges by our choice of parameters, while some scenario A points are excluded, and we use open markers to denote them.
}
\label{fig:MeePlot}
\end{figure}

\begin{figure}[h!]
\centering
\includegraphics[width=0.9\textwidth]{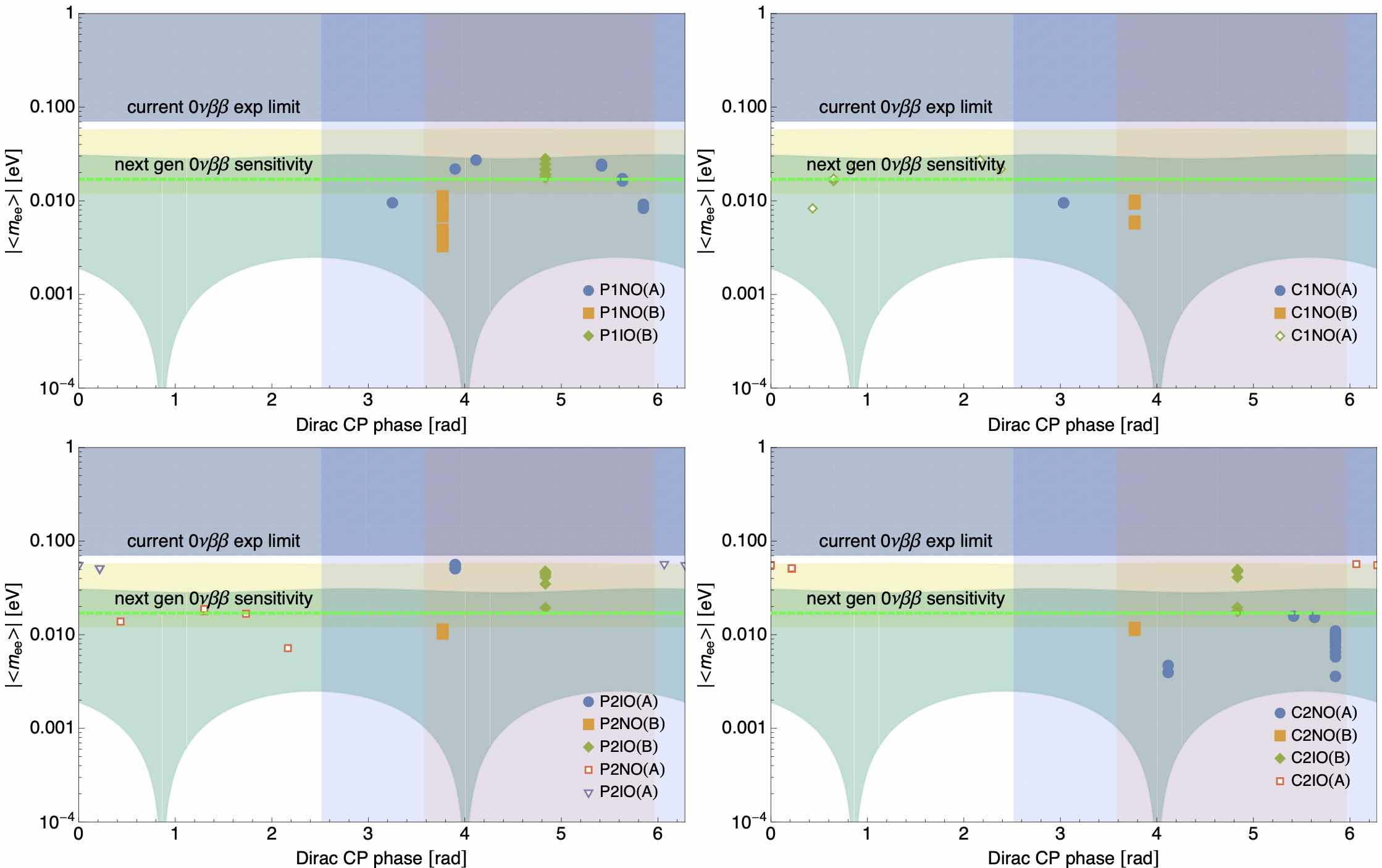}
\includegraphics[width=0.9\textwidth]{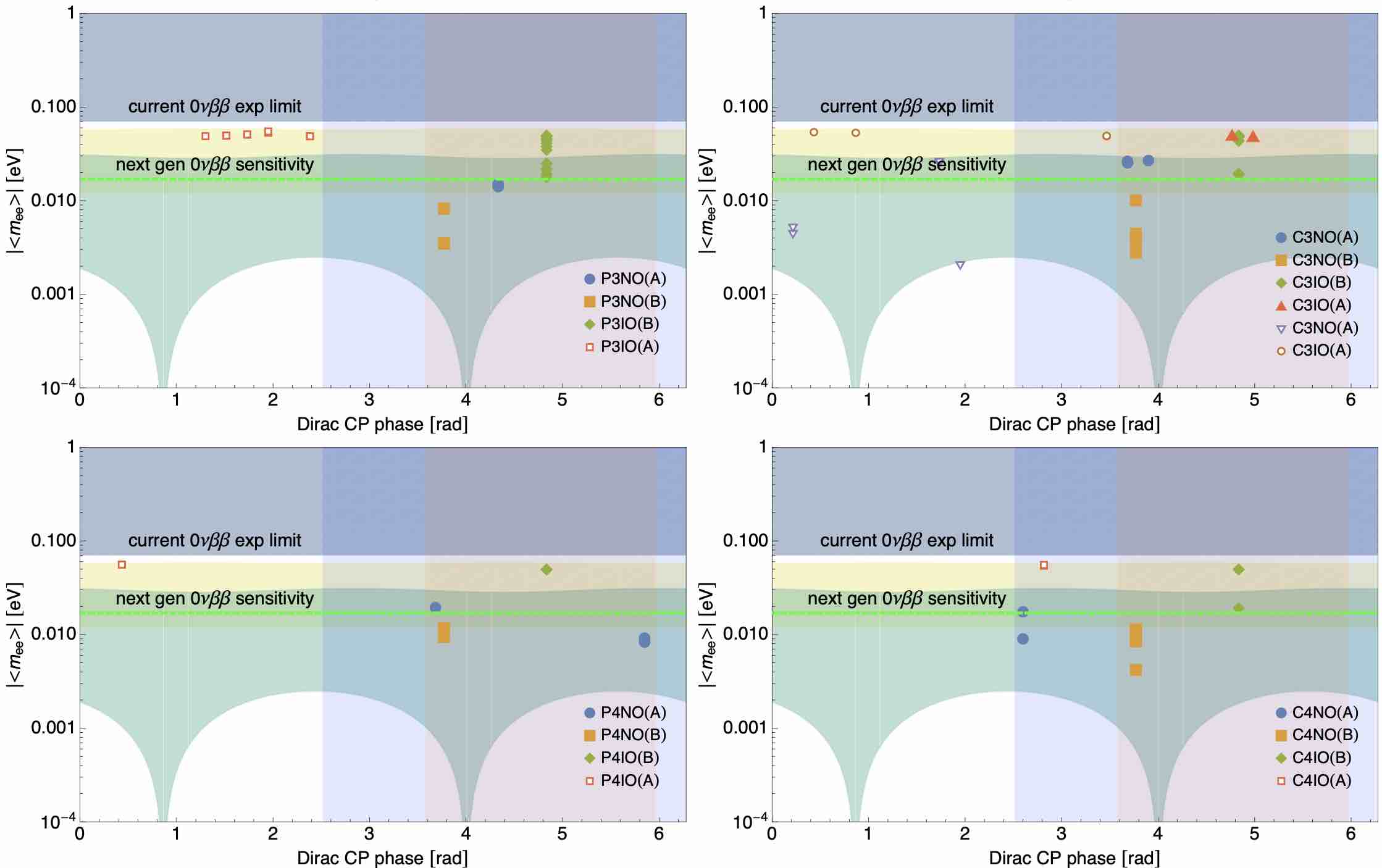}
\caption{The effective Majorana mass with dependence on the Dirac CP phase. The basic settings are the same as Fig.~\ref{fig:MeePlot}. We show the $3\sigma$ region of the Dirac CP phase as a shaded blue (orange) area for normal (inverted) ordering.
}
\label{fig:MeedPlot}
\end{figure}

\begin{table}[t!]
 \caption{\label{tab:sum} Summary of BAU's dependence on low-energy parameters.}\vspace{0.12cm}
 \centering
  \begin{tabular}{c|c|c|c|c}
   \toprule \hline
   & \multicolumn{3}{c|}{ Dirac phase only} & Majorana phases added-on \\ \hline
 cases & lightest neutrino mass [eV]\footnotemark & $Y_\mathrm{B}$ manitude & $Y_\mathrm{B}$ sign & $Y_\mathrm{B}$ magnitude \\ \hline
P1NO &  $[0.004,0.03]$ & $\checkmark$ &  $\checkmark$ & $\checkmark$ \\
P1IO &   -- & \xmark  & $\checkmark$  & $\checkmark$ \\ 
C1NO &  $[0.018,0.03]$  & $\checkmark$   & \xmark   & $\checkmark$ \\ 
C1IO &   -- & \xmark   & \xmark & \xmark  \\ 
&&&&\\
P2NO &  --  & $\checkmark$   & \xmark   & $\checkmark$ \\ 
P2IO &  --   & $\checkmark$ &  $\checkmark$ & $\checkmark$ \\
C2NO &  $[10^{-4},0.03]$    & $\checkmark$ &  $\checkmark$ & $\checkmark$ \\
C2IO &    -- & $\checkmark$ &  \xmark  & $\checkmark$ \\
&&&&\\
P3NO &  $[10^{-4},0.03]$      & $\checkmark$ &  $\checkmark$ & $\checkmark$ \\
P3IO &   --    & $\checkmark$ &  \xmark  & $\checkmark$ \\
C3NO & $[10^{-4},0.03]$       & $\checkmark$ &  \xmark  & $\checkmark$ \\
C3IO  &   $[10^{-4},0.03]$     & $\checkmark$ &  $\checkmark$ & $\checkmark$ \\
&&&&\\
P4NO &  $[10^{-4},0.03]$       & $\checkmark$ &  $\checkmark$ & $\checkmark$ \\
P4IO  &   --    & $\checkmark$ &  \xmark  & $\checkmark$ \\
C4NO &   $[0.023,0.03]$     & $\checkmark$ &  \xmark  & $\checkmark$ \\
C4IO  &     --  & $\checkmark$ &  $\checkmark$ & $\checkmark$ \\
   \hline
  \bottomrule
\end{tabular}
\end{table}
\footnotetext{This column shows the lightest neutrino mass range given by the observed $Y_\mathrm{B}$ with correct magnitude and sign. The ``$Y_\mathrm{B}$ magnitude" column means it is sufficient to give a correct $Y_\mathrm{B}$ magnitude (but in many cases much larger). For example, case C4IO gives $Y_\mathrm{B}$ much larger than observed value, so no constraint is given for the lightest neutrino mass.}

Low energy CP violation, represented by the CP-violating phases contained in the lepton mixing matrix, is crucial to leptogenesis. In MLRSM, with unbroken LR symmetry in the lepton sector, the low-energy CP violation is the only source of CP violation needed in leptogenesis. A direct link between low-energy CP violation and BAU is established in such circumstances. Future discovery of the low-energy CP-violating phases will lead to direct estimation of BAU. There is strong evidence that the Dirac phase is non-zero. The global fitted Dirac phase value is even close to its maximal value. Long baseline oscillation experiments are aiming to get a precise measurement of this phase in ten years. This makes the investigation of leptogenesis generated by this phase necessary and timely.

Our numerical results show in many cases, the observed baryon asymmetry can be generated with CP violation sources are totally in the low-energy neutrino sector, i.e., the Dirac CP phase and the two Majorana CP phases. The results are summarized in Table~\ref{tab:sum}. 
From Table~\ref{tab:sum}, we see that the Dirac phase can be the only source of CP violation and lead to successful leptogenesis in cases P1NO, P2IO, C2NO, P3NO, C3IO, P4NO, and C4IO. By successful, we mean both a correct magnitude and a correct sign of BAU. When Majorana phases are added to play, all cases except C1IO can accommodate the desired amount of BAU.

The oscillation experiments are insensitive to Majorana phases, while these phases contribute to the lepton asymmetry in leptogenesis. These phases could be probed in the neutrinoless double beta decay ($0\nu\beta\beta$) experiments. The half-life $T_{0\nu}^{1/2}$ measured by the $0\nu\beta\beta$ experiments has a dependence on neutrino masses and mixings through an effective neutrino mass, which is defined as
\begin{align}
|\langle m_{ee} \rangle| = | \sum_{i=1}^3 U_{ei}^2 m_i|.
\end{align}
All channels involving $W_\mathrm{R}$ and other mediators like heavy neutrinos or triplet scalars in MLRSM are highly suppressed due to the scale of current model, thus can be safely neglected.

From our scanned data set, we select the points that lie in $3\sigma$ region of the observed $Y_\mathrm{B}$ and plot them in Fig.~\ref{fig:MeePlot} and Fig.~\ref{fig:MeedPlot}. In these plots, we fill the points in agreement with current oscillation data and use open markers for the rest. From both plots, we see that in several cases, there are points within next-generation $0\nu\beta\beta$ experiments' reach, and corresponding parameter space can be probed by then. It is noticeable that next-generation neutrinoless double beta decay experiments, together with the oscillation experiments, can probe the parameter space given by current $16$ cases to verify or falsify these models.
Note that the total number of the points that lie in $3\sigma$ region of the observed $Y_\mathrm{B}$ selected from our scattered plots is not large, so we cannot draw any conclusions about the upper and lower limit on the resulting effective neutrino mass.

\section{Conclusions}\label{sec:conclusion}

We perform a thermal unflavored leptogenesis analysis on MLRSM models. With unbroken LR symmetry in the lepton Yukawa sector, the neutrino Dirac coupling is totally determined by the light and heavy neutrino information, leaving no orthogonal matrix ambiguity. We further choose to work with $V_\mathrm{R}=V_\mathrm{L}^*$ for $\mathcal{C}$ being the LR symmetry ($V_\mathrm{R}= V_\mathrm{L}$ for $\mathcal{P}$ being the LR symmetry), leaving CP violation all resides in the low-energy sector. Both type I and mixed type I$+$II neutrino mass generation mechanisms are considered, together with different CP asymmetry generation mechanisms with either heavy neutrino or left-handed triplet decay. Considering either $\mathcal{C}$ or $\mathcal{P}$ can be the LR symmetry and both mass orderings of light neutrinos, we obtain sixteen cases. In many cases, the observed baryon asymmetry can be generated with CP violation totally in the low-energy neutrino sector, i.e., the Dirac CP phase and the two Majorana CP phases.

Future long-baseline experiments like T2K, NO$\nu$A, DUNE, T2HK, and T2HKK will narrow down the Dirac phase range, which allows us to distinguish among the sixteen cases when only the Dirac phase contributes. In cases with Majorana phases contribution included, we show that several cases predict an effective Majorana mass within next-generation neutrinoless double beta decay experiments' reach.

Our findings show that in MLRSM, when  $\mathcal{C}$ or $\mathcal{P}$ is unbroken in the lepton Yukawa sector, the low-energy CP phases in the lepton mixing matrix are sufficient to reproduce the observed BAU.  It establishes the connection between the low-energy CP violation and the baryon asymmetry of the Universe, with neither ambiguity in the arbitrary orthogonal matrix nor model-dependent assumptions about the neutrino Dirac coupling. The models' parameter space can be probed by current and future oscillation and neutrinoless double beta decay experiments.

With fixed heavy sector parameters, we obtain how the final baryon asymmetry depends on low-energy parameters. Given our positive results, one can further investigate the flavored regime with lower or varying heavy scales, or an unconstrained heavy neutrino mixing, which is expected to be interesting, and we leave it for a future study. On the other hand, we work with a particular case that light and heavy neutrino mixing coincidence and leave a generic treatment in a future work.

\section*{Acknowledgement} 

JHY would like to thank Hao-Lin Li and Kaori Fuyuto for the discussions and collaboration in the early stage of this project. We would like to appreciate Juan Carlos Vasquez Carmona for discussions and his valuable comments on the manuscript. JHY is supported by the National Science Foundation of China (NSFC) under Grants No. 11875003 and No. 11947302.  X. Z. is supported by China Postdoctoral Science Foundation under Grant No. 2019M650001.

\appendix
\section*{Appendices}\label{sec:app}
\addcontentsline{toc}{section}{Appendices}
\renewcommand{\thesubsection}{\Alph{subsection}}

\subsection{Type II mass domination, $v_\mathrm{L}/v_\mathrm{R}$ and heavy neutrino spectrum}\label{sec:appdx1}

\begin{figure}[t!]
\centering
\includegraphics[width=.8\textwidth]{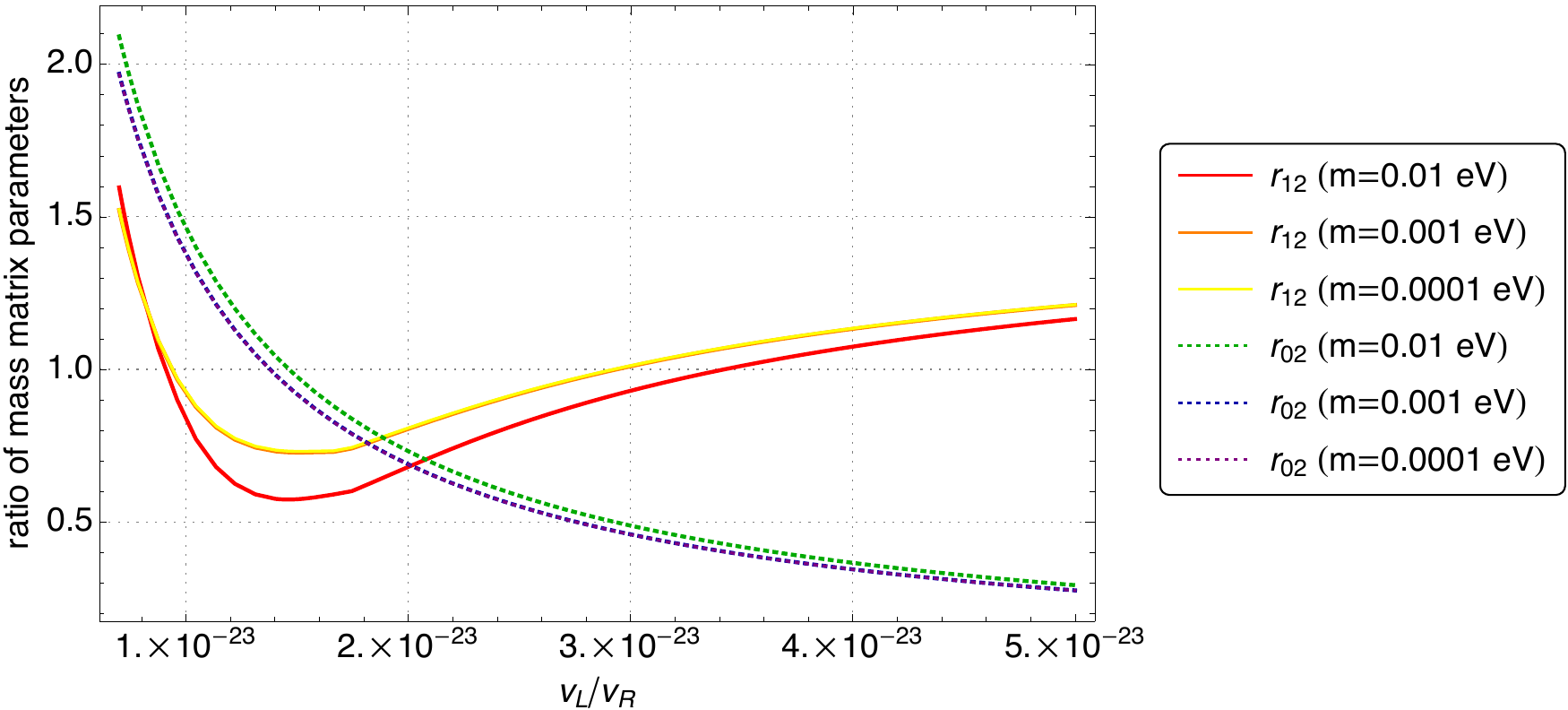}
\caption{The ratios of the mass matrix parameters as functions of $v_\mathrm{L}/v_\mathrm{R}$.
}
\label{fig:rPlot}
\end{figure}

To quantify the contribution to the light neutrino mass, we introduce the following dimensional parameters
\begin{align}
\overbar{M_\nu} &\equiv \sqrt{\mathrm{Tr}\left( M_\nu^{\dagger} M_\nu \right)},\\
\overbar{M_\nu^\mathrm{I}} &\equiv \sqrt{\mathrm{Tr}\left( M_\nu^{\mathrm{I}\dagger} M_\nu^{\mathrm{I}} \right)},\\
\overbar{M_\nu^\mathrm{II}} &\equiv \sqrt{\mathrm{Tr}\left( M_\nu^{\mathrm{II}\dagger} M_\nu^\mathrm{II} \right)},
\end{align}
and use them to compare the ``relative magnitude" of two matrices. In type I mass dominating cases, we require that $\overbar{M_\nu^\mathrm{I}}/\overbar{M_\nu^\mathrm{II}} \gtrsim 10^2$. 
%and is shown in the same subsection \ref{subsec:overview}. Note that $\overbar{M_\nu}$ is fully determined once the lightest light neutrino mass is given, so it can serve well as a measure, which is what we did in type II mass domination cases. 
In type II mass dominating cases, there is no such a simple definition. Both $M_\nu^\mathrm{I}$ and $M_\nu^\mathrm{II}$ depend on $v_\mathrm{L}/v_\mathrm{R}$ ($M_\nu^\mathrm{I}$ has $v_\mathrm{L}/v_\mathrm{R}$ dependence through $M_\mathrm{D}$), so do $\overbar{M_\nu^\mathrm{I}} $ and $\overbar{M_\nu^\mathrm{II}}$. As a result, $\overbar{M_\nu^\mathrm{I}}$ cannot be tuned arbitrarily small.  We take the P2NO case as an example to illustrate the issue. In Fig.~\ref{fig:rPlot}, we plot the following mass contribution ratios as a function of $v_\mathrm{L}/v_\mathrm{R}$
\begin{align}
r_{02} &= \overbar{M_\nu} /\overbar{M_\nu^\mathrm{II}} ,\\
r_{12} &= \overbar{M_\nu^\mathrm{I}} /\overbar{M_\nu^\mathrm{II}}.~\label{eq:rij}
\end{align}
The optimistic $v_\mathrm{L}/v_\mathrm{R}$ value is that gives either $r_{02} $ approach $1$ from above, or is at $r_{12}$'s minimum. They both point to $v_\mathrm{L}/v_\mathrm{R}\simeq 1.4\times 10^{-23}$. We fix it to $v_\mathrm{L}/v_\mathrm{R} = 1.45\times 10^{-23}$, which gives $r_{02}=1.01$ and $r_{12}=0.57$. This would come with no surprising because $\overbar{M_\nu} \neq \overbar{M_\nu^\mathrm{I}} +\overbar{M_\nu^\mathrm{II}} $. We see from this example that the type II mass dominating case is actually a mixed type I$+$II with type II mass dominance.  

In type II mass dominating cases, we cannot get a pure type II mass domination for the following reason. To encode the neutrino Dirac coupling $M_\mathrm{D}$ in terms of the light and heavy neutrino information, we cannot neglect the type I mass which contains $M_\mathrm{D}$. Because with only the type II mass contribution, one cannot establish a connection of $M_\mathrm{D}$ (which only shows up in type I contribution) with the light neutrino mass matrix, and thus cannot get rid of the ambiguity caused by the Casas-Ibarra $R$ matrix. So working in our setup (takeing $M_\mathrm{D}$ from Eq.(\ref{eq:MD_C}) or Eq.(\ref{eq:MDH}) ) means that the type I term always show up. Note that $M_\mathrm{D}$ has $v_\mathrm{L}/v_\mathrm{R}$ dependence. So the type I mass also has $v_\mathrm{L}/v_\mathrm{R}$ dependence. As a result, the type I mass contribution cannot be tuned arbitrarily small, and the mass contribution ratios (the $r_{ij}$ parameters introduced in Eq.(\ref{eq:rij})) always have a shape like Fig.~\ref{fig:rPlot}.  

Since the light and heavy neutrino mass are proportional to each other only when the type I mass is omitted and is not our case here, they do not have the same mass ordering. So it is acceptable to fix the heavy neutrino mass spectrum in our type II mass domination case.

\subsection{Rate densities}

\subsubsection{Heavy neutrino decay}
The two body decay rate density is
\begin{align}
\gamma_\mathrm{D} (z) = \displaystyle\frac{T^3}{\pi^2} z^2 K_1(z) \Gamma_\mathrm{D},
\end{align}
where $K_1$ is a Bessel function and $\Gamma_\mathrm{D}$ is the total decay width.
For scattering process, the rate density is 
\begin{align}
\gamma (a + b\rightarrow i + j + ...) =\displaystyle \frac{T}{64 \pi^4} \int_{(m_a + m_b)^2}^\infty ds~ \hat{\sigma}(s) \sqrt{s} K_1(\frac{\sqrt{s}}{T}),
\end{align}
where $\hat{\sigma}$ is the reduced cross section which relates to the usual cross section as 
$$\displaystyle \hat{\sigma} (s) =8\left[ (p_a \cdot p_b)^2-m_a^2 m_b^2\right] \sigma (s)/s.$$

To show the various reduced cross sections, we follow the notation in Ref.~\cite{Giudice:2003jh} and start with 
\begin{align}
z &= \frac{m_{N_1}}{T},~\displaystyle x=\frac{s}{m_{N_1}^2},  ~\displaystyle a_{H,L,Q,U,W,B} = \frac{m^2_{H,L,Q,U,W,B}}{m^2_{N_1}}, \nonumber\\
\displaystyle a_\Gamma &= \frac{\Gamma^2_{N_1}}{m^2_{N_1}}, ~D_{N_1}=\frac{1}{x-1+i a_\Gamma^{1/2}}, ~|D_{N_1}^2|^{\rm sub}=\frac{(x-1)^2-a_\Gamma}{\left[(x-1)^2+a_\Gamma\right]^2}.
\end{align}
The $LN \leftrightarrow Q_3 U_3$ reduced cross section is
\begin{align}
\displaystyle  \hat{\sigma}_{Hs} = \frac{3}{4\pi} \left(Y_N^{} Y_N^\dagger\right)_{11} y_t^2 \frac{(x-1-a_L)(x-2a_Q)}{x(x-a_H)} \sqrt{\left[ (1+a_L-x)^2-4a_L\right] (1-4a_Q/x)}.
\end{align}
The $\overline{U}_3 N \leftrightarrow Q_3 \overline{L}$ and the  $\overline{Q}_3 N \leftrightarrow U_3 \overline{L}$ cross sections are 
\begin{align}
\hat{\sigma}_{Ht} =& \frac{3}{4\pi} \left(Y_N^{} Y_N^\dagger\right)_{11}   y_t^2 \frac{1}{x} \nonumber\\
&\left[  t_+ - t_- -(1-a_H+a_L)(a_H -2a_Q) \left( \frac{1}{a_H-t_+}  -  \frac{1}{a_H-t_-}\right) \right. \nonumber\\
&\left. -(1-2a_H+a_L+2a_Q) \mathrm{ln}\frac{t_+-a_H}{t_- -a_H}  \right],
\end{align}
where 
\begin{align}
t_\pm = &\frac{1}{2x} [ a_Q + x - (a_Q - x)^2 + a_L (x + a_Q -1) \nonumber\\
& \pm \sqrt{ \left[ a_Q^2 + (x-1)^2 -2a_Q (1+x)\right] \left[ a_L^2 + (x-a_Q)^2 -2a_L (a_Q+x)\right]}.
\end{align}
For $\Delta \mathrm{L} =1$ scattering involving gauge bosons, the $L N_1 \rightarrow \overline{H} A$ cross section is
\begin{align}
\hat{\sigma}_{As} = &\displaystyle\frac{3g_2^2}{16 \pi x^2}  \left(Y_N^{} Y_N^\dagger\right)_{11} \left[ 2t(x-2) +(2-2x+x^2) \mathrm{ln}[(a_L-t)^2+ \epsilon] \right.\nonumber\\
&\left. + 2 \frac{ x(a_L - t) (a_L + a_L x -a_W)+\epsilon (2-2x+x^2)}{(a_L-t)^2+\epsilon} \right]^{t_+}_{t_-}.  ~\label{eq:HAs}
\end{align}
The $\overline{L} A \rightarrow N_1 H$ and the $\overline{L} \overline{H} \rightarrow N_1 A$ cross section is
\begin{align}
\hat{\sigma}_{At} = & \displaystyle\frac{3 g_2^2}{32 \pi x}  \left(Y_N^{} Y_N^\dagger\right)_{11}  \left[\frac{2}{1-x} \left[ 2x \mathrm{ln} (t- a_H) - (1+x^2) \mathrm{ln} (t+x-1-a_W-a_H)\right]\right.\nonumber\\
& \left.+\frac{1}{x}  \left[ t^2+ 2t(x-2) -4(x-1) \mathrm{ln}(t-a_H) +x \frac{a_W-4a_H}{a_H-t} \right]   \right]^{t_+}_{t_-} .
\end{align}
Here and in Eq.~(\ref{eq:HAs}) 
\begin{align}
t_\pm = \displaystyle\frac{(m_1^2-m_2^2-m_3^2+m_4^2)^2}{4s} - \left(  \sqrt{\frac{(s+m_1^2-m_2^2)^2}{4s} -m_1^2}  \pm \sqrt{\frac{(s+m_3^2-m_4^2)^2}{4s} -m_3^2}  \right)^2.
\end{align}

\subsubsection{Left-handed triplet scalar decay}
The decay rate density is 
\begin{align}
\gamma_\mathrm{D} (z) = s(z) \Gamma_\Delta \Sigma_\Delta^{eq}(z) \displaystyle\frac{K_1(z)}{K_2(z)}.
\end{align}
For ``$\Delta \mathrm{T} = 2$" scattering involving gauge bosons, the reduced cross section is~\cite{Hambye:2005tk,Hambye:2012fh}
\begin{align}
\hat{\sigma}_A =\frac{6}{72\pi} \left\{ (15 C_1 - 3 C_2) r + (5C_2 - 11C_1) r^3 + 3(r^2-1) \left[2C_1 + C_2 (r^2-1)\right] \mathrm{ln}\frac{1+r}{1-r} \right\},
\end{align}
where $x=s/m_\Delta^2, r=\sqrt{1-4/x}, C_1=3g^4/2+ 3g^4_Y + 12g^2 g^2_Y, C_2=6g^4+ 3g^4_Y + 12g^2 g^2_Y$.\\
For $\Delta \mathrm{L} =2$ process, we adopt~\cite{Lavignac:2015gpa}
\begin{align}
\hat{\sigma}_{lH} = \displaystyle \frac{3x m_\Delta^2}{2 \pi v^2} \sqrt{\mathrm{Tr}\left(M_\nu^\mathrm{I}M_\nu^{\mathrm{I}\dagger}\right)}
\end{align}
for type I mass domination. While for mixed type I + II with type II domination, we use
\begin{align}
\hat{\sigma}_{lH} = \displaystyle \frac{3x m_\Delta^2}{4 \pi v^2}\mathrm{Re}\left[ \mathrm{Tr}\left(M_\nu^\mathrm{II}M_\nu^{\mathrm{I} \dagger}\right) \right] \left\{ \frac{1-x}{(1-x)^2+\epsilon^2} + 2\left[1-\frac{\mathrm{ln}(1+x)}{x}\right]\right\}.
\end{align}

\subsection{Effects of heavy neutrino mass spectrum}\label{apd:spectrum}

\begin{figure}[t!]
\centering
\includegraphics[width=0.9\textwidth]{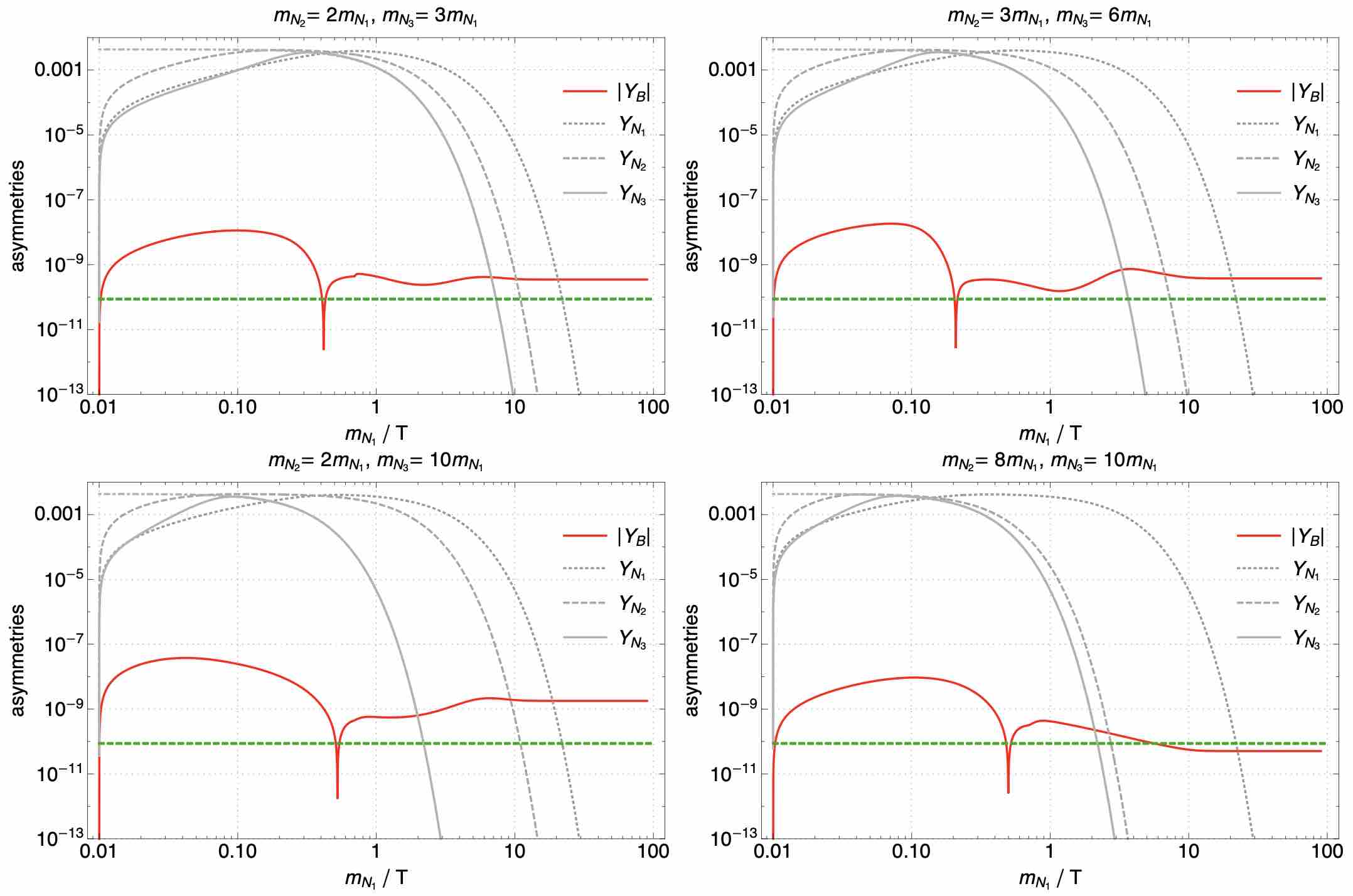}
\caption{ The baryon asymmetry and heavy neutrino comoving number densities as a function of $z=m_{N_1}/T$ in the case P1NO for different heavy neutrino mass spectra. The horizontal green dashed lines are the $3\sigma$ region of observed baryon asymmetry $Y_\mathrm{B}=(8.72\pm 0.08)\times 10^{-11}$~\cite{Aghanim:2018eyx}. The oscillation parameters are fixed at their best fit values, and the Majorana phases are set to zeros. $m_1$ is fixed to $0.01$ eV.
}
\label{fig:Nspectrum}
\end{figure}

As we are interested in the role of the CP-violating phases, we fix the heavy neutrino mass spectrum to be $m_{N_2} = 2 m_{N_1},~m_{N_3} = 3 m_{N_1}$ before in the main contexts. This choice corresponds to a ``slice" of the parameter space, which allows us to focus on the role of the low-energy CP-violating phases. %An investigation of the whole parameter space is very involved, if not impossible. If the heavy neutrino mass spectrum changes, other parameters will change accordingly to have a correct amount of BAU. However, our main conclusion stays the same: the low-energy CP violation can be the only source of CP violation need in generating BAU through leptogenesis in MLRSM with unbroken left-right symmetry.

%The viability and testability of 16 cases are subject to the whole parameter space which include the fixed mass parameters. With these benchmark points, we have shown that a connection between low-energy CP violation and BAU without ambiguity caused by the arbitrary orthogonal R matrix can be established. We are positive that for other values of the benchmark points, similar connection can be established in different parameter space, and that is the meaning of the study, i.e., to give a viable example.

It is interesting to check the effects of the heavy neutrino mass spectrum with a few examples in Fig.~\ref{fig:Nspectrum}, where we show the baryon asymmetry and heavy neutrino comoving number densities as a function of $z=m_{N_1}/T$ for different heavy neutrino mass spectra. The top-left plot is the same as the right plot of Fig.~\ref{fig:ratePlotp1no}, and the other three are used for comparison. We see that the first three all have a BAU larger than needed. Considering the CP-violating phases and the lightest neutrino mass may reverse the sign of the CP asymmetries, there should be no problem finding the correct amount of BAU in these cases, as shown for the first one in the main contexts. The same thing would happen for the last case considering that the Majorana phases provide additional CP-violating sources. 

Looking at the heavier neutrino comoving densities $Y_{N_2},~Y_{N_3}$ (grey dashed and grey lines in Fig.~\ref{fig:Nspectrum}), we find that the heavier their masses, the earlier their densities vanish. As a result, for a very hierarchical mass spectrum, it is usually sufficient to consider only one heavy neutrino decay.

An investigation of the whole parameter space is very involved, if not impossible. If the heavy neutrino mass spectrum changes, other parameters will change accordingly to have the correct amount of BAU. However, our main conclusion stays the same: the low-energy CP violation can be the only source of CP violation need in generating BAU through leptogenesis in MLRSM with unbroken LR symmetry.

\end{document}